\documentclass[epj]{svjour}
\usepackage{latexsym}
\usepackage{amsmath,amscd}
\usepackage{amsfonts,amssymb}
\usepackage{epsfig}
\usepackage{graphics}
\usepackage{rawfonts}
\newcommand{\npar}{\bigskip\par\noindent}

\newcommand{\beq}[1]{\begin{equation}\label{#1}}
\newcommand{\eeq}{\end{equation}}
\newcommand{\bmath}{$$}  \newcommand{\emath}{$$}
\newcommand{\bear}[1]{\begin{eqnarray}\label{#1}}
\newcommand{\eear}{\end{eqnarray}}

\newcommand{\lit}[3]{{~\bf #2},~#3~(#1)}  % Phys.Rev., EPJ

\def\etal {{\it et al.}}
\def\coll {{\rm Coll.}}

\def\EJP {{Eur. J. Phys.}}
\def\NIM {{Nucl. Instr. Meth.}}
\def\PRL {{Phys. Rev. Lett.}}

\def\PRD {{Phys. Rev. D}}
\def\PL {{Phys. Lett.}}

\def\NP {{Nucl. Phys.}}

\begin{document}

\title{Study of the Jet Shape at 920 GeV/c
in Proton-Nucleus Interactions with HERA-B Detector\\
}

\author{D.~Golubkov\inst{1}\thanks{\emph{e-mail:} dimgol@mail.desy.de}%
\and  Yu.~Golubkov\inst{2}\thanks{\emph{e-mail:} golubkov@mail.desy.de}%
}

\institute{Moscow Physical Engineering Institute, 115409 Moscow, Russia
\and Institute of Nuclear Physics of Moscow State University, 119899 Moscow, Russia}

\date{Received: date / Revised version: date}
% The correct dates will be entered by Springer
%

% - - - - - - - - - - - - - - - - - - - - - - - - - - -
\abstract{We performed a measurement of differential and integral jet shapes
in proton-carbon, proton-tungsten 
and proton-aluminium collisions at 920~GeV/c proton momentum 
with the HERA-B detector at HERA for the jet transverse energies
in the range $4\,<\,E_T(jet)\,<\,12$~GeV. Jets were identified
using the $k_T$-clustering algorithm. The measurements were performed
for the hardest jet in the event, directed towards the opposite side
with respect to the trigger direction. Jets become narrower with
increasing transverse energy and measured distributions
agree well with predictions of the PYTHIA~6.2 model.
We do not observe any significant difference in the jet shape 
for the carbon and the aluminium targets.
Nevertheless, the transverse energy flow at small and large radii 
for the tungsten sample is slightly less than for light nuclei. 
This observation indicates some influence of the nuclear environment
on the formation of jets in heavy nuclei, especially at lower transverse 
energies, $5\,<\,E_T(jet)\,<\,6$~GeV.
 % end of PACS codes
} %end of abstract

\authorrunning{D.~Golubkov, Yu.~Golubkov}
\titlerunning{Study of Jet Shape at 920 GeV/c}

\maketitle

% - - - - - - - - - - - - - - - - - - - - - - - - - - -
\bigskip
\section{Introduction}
\label{sec:intro}
The parton$\,\to\,$hadron transition is one of the most interesting
outstanding questions in QCD and we should look inside jets to better
understand this process.

In the Standard Model our general understanding of high-energy collisions 
of hadrons suggests that jets arise when short-dis\-tance, 
large-\-mo\-men\-tum-tran\-sfer
interactions generate partons (quarks and gluons) that are widely separated
in momentum space just after the hard collision. In a fashion that is not
yet quantitatively understood in detail these configurations are thought
to evolve into hadronic final states exhibiting collimated sprays of hadrons,
which are called jets. Thus jets can be
regarded as a universal signal of parton dynamics at short distances.

Hadron collisions are a perfect place to perform such studies, because
they are a high rate source of jets over a very wide range
of QCD scales in the same experiment.
High-$p_T$~ jets dominate the event structure in hadronic collisions
at high center-of-mass (CM) energies $\sqrt{s}$~ when sufficiently large
transverse energy $E_T$~ is required. This was demonstrated by various
experiments at $\sqrt{s}\,=\,63$~GeV \cite{Akesson83}, 
at $\sqrt{s}\,=\,540$~GeV
\cite{Banner82}, at $\sqrt{s}\,=\,1800$~GeV \cite{Abe89} at hadronic
colliders.

At large transverse energies ($E_T\,\sim\,100$~GeV) the corrections to jet
shape from parton fragmentation are usually considered to be small
in comparison to ones due to the parton cascade initiated by the high-$E_T$~
scattered parton. But at the more moderate transverse energies $\sim$~ 
of a few GeV,
this conclusion is not valid. Thus the investigation of the jet structure
at moderate energies in hadronic collisions can give important
information about parton fragmentation processes in hadronic interactions.

Measurements of jets in fixed-target experiments  not only widen
the energy range of jet studies but, in addition, enable extension of these 
studies into a new realm of colliding particles: meson-nucleon and
hadron-nucleus interactions. 

At lower $\sqrt{s}$, because of contributions from mechanisms such
as initial- and final-state parton radiation and multiple scattering
of quarks and gluons, the event structure rarely exhibits dijet topology 
and the jet signal is rather difficult to extract experimentally
\cite{Marzo83,Brown84}. Nevertheless,  the presence of jet structure
has also been demonstrated in fixed target experiments 
at lower energies specifically at $800$~GeV in $pA$-collisions 
\cite{Stewart90}, 
and at $500$~GeV in both $pBe$~ and $\pi Be$-collisions \cite{Alverson94}.
Also a jet signal consistent with QCD predictions and with
extrapolations of jet cross sections from higher CM energies was found
by \cite{Cornell85,Arenton85}.

The methods employed to extract jets at moderate 
CM energies depend heavily on Monte Carlo models. Therefore, the absolute
cross sections for jet production determined at these energies are subject
to large systematic errors. However, the relative dependence of jet production
and properties from different nuclear targets should be less sensitive to
the assumed jet size and background.

There is a large variety of jet-shape variables, 
they are very informative and enter all hard processes with features 
which are expected to be universal (QCD factorization).
In the present analysis we concentrate on studies of the internal 
jet structure,
measuring the differential and integral jet shape.

% ==============================================================

\section{Jet definition}
\label{sec:jetdef}

A jet is qualitatively defined as a collimated spray of high-energy hadrons.
However, for the purpose of performing accurate quantitative studies,
one needs a precise definition of a jet. Essentially, one has to specify how
low-energy particles are assigned to jets, in order to have 
infrared-finite cross sections.

The standard ``Snowmass convention'' on jet 3-momentum definition 
\cite{Jetdef} is:

\beq{jetcomb}
\begin{array}{lll}
E_T(jet) & = &\sum_i\,p_{T,i}\,, \\
\\
\eta(jet) & = & \sum_i\,p_{T,i}\,\eta_i\,/E_T(jet)\,, \\
\\
\phi(jet) & = & \sum_i\,p_{T,i}\,\phi_i\,/E_T(jet)\,,
\end{array}
\eeq

\npar with $E_T(jet),\,\eta(jet),\,\phi(jet)$~ being the jet transverse energy,
pseudo-rapidity and azimuthal angle, respectively, and
$p_{T,i},\,\eta_i,\,\phi_i$~ are transverse energy, pseudo-rapidity 
and azimuthal angle of particles, forming the jet. Usage of the above 
observables ensures the invariance of the jet momentum definition 
with respect to longitudinal Lorentz boosts.

\bigskip
Beside the above general definition of the jet's characteristics one needs
to specify an algorithm -- how to assign a particle to one of the jets
in the event. 
The most usable at present are two algorithms -- ``cone'' and 
``$k_T$'' (or ``Durham'') algorithms. 

The simplest cone algorithm (see, e.g., \cite{Jetdef}) starts from
a ``jet initiator'' -- a particle with transverse
momentum above a predefined threshold, e.g., $p_T\,\ge\,1$~GeV, 
adjoining to the initiator particles 
within a cone of radius $r\,\le\,R_c$.
Here $r\,=\,\sqrt{\Delta\eta^2\,+\,\Delta\phi^2}$~ and $R_c\,\approx\,0.7$.
There are modifications of this algorithm, using an iterative procedure 
to define the current jet axis.
The cone algorithm is suitable for $e^+e^-$~ collisions, where the remnants
of the initial particles are absent.
For hadronic interactions and deep-ine\-las\-tic lep\-ton-had\-ron 
scat\-te\-ring, 
the cone algorithm meets problems with particles 
separation in multi-jet events as well problems when being compared 
with NLO and NNLO QCD calculations.

On the other hand the $k_T$ algorithm \cite{Soper93,Catani93} is expected 
to be the best for hadronic collisions because it is based on the QCD picture 
of jet development. This algorithm is invariant under longitudinal Lorentz 
boosts and is infrared and collinear safe.
It has been shown \cite{Ellis92} that the inclusive $k_T$ cluster
algorithm provides, at present, the best jet finding algorithm 
from the theoretical 
point of view, since the problem of overlapping jets, which affects, e.g.,
the iterative cone algorithm \cite{Abe92}, is avoided.

The measure of the ``closeness'' of two particles/protojets 
in the $k_T$ algorithm  is:

\beq{measure}
\begin{array}{lll}
d_{ij} & = & \min \left (p_{T,i}^2,p_{T,j}^2\right )\\
 & \times\, &
\left [\left (\eta_i-\eta_j\right )^2+
\left (\phi_i-\phi_j\right )^2\right ]\,.
\end{array}
\eeq

We use here the original expression for $d_{ij}$~ \cite{Soper93}
without the additional parameter $R_0$~ suggested in \cite{Seymour97},
where the measure (\ref{measure}) is replaced by $d_{ij}/R_0^2$.
Because the parameter $R_0$~ is to be around unity, it does not make
sense to introduce $R_0$~ in our case.
At small angles between two particles 
the measure, $d_{ij}$, is, approximately, the squared relative transverse
momentum of one of the particles with respect to the other particle. 

The $k_T$ algorithm looks as following \cite{Soper93}:
% - - - - - - - - - - - - - - - - - - - - - - - - - - 
\bigskip
\begin{enumerate}
\item For each protojet, define $d_i\ =\ E_{T,i}^2$
and for each pair of protojets define $d_{ij}$ from (\ref{measure})
\item Find the smallest of all the $d_i$ and $d_{ij}$ and label it $d_{min}$
\item If $d_{min}$ is a $d_{ij}$, merge protojets $i$ and $j$ into a new
protojet $k$ according to (\ref{jetcomb})
\item If $d_{min}$ is a $d_i$, the corresponding protojet $i$ 
is ``not mergeable''. 
Remove it from the list of protojets and add it to the list of jets.
\end{enumerate}

Repeating the above four steps till there are ``mergeable'' protojets
one subdivides the event into a number of groups of particles
where each particle is assigned to one and only one group/jet.

% ==============================================================
\section{Jet shape observables}
\label{sec:jetobs}

In the present work we measure the differential $\rho(r)$ \\
and integral $\Psi(r)$~ jet shapes characterizing 
how widely a jet's energy is spread in the $(\eta,\phi)$~plane.
Jet shape is one of the most popular characteristics of jet structure 
in hadron collisions. 
Jet shape §$\Psi(r)$~ is defined as the transverse
energy flow within the cone of radius $r$~ around the jet axis
in $(\eta,\phi)$~plane normalized to unity and averaged over all jets 
in the (sub)sample.
The differential jet shape $\rho(r)\,dr$~ is the derivative of $\Psi(r)$~
over $r$~ and is the transverse energy flow through the annulus of width
$dr$~ with radius $r$~ around the jet axis. 

These observables are, in general, collinear safe (for two particles 
with parallel momenta $\vec p_i, \,\vec p_j$~ one may replace 
these two particles with a single one with a momentum
equal to $\vec p_i\,+\,\vec p_j$) and infra-red safe (for $p_i\,\ll\,p_j$~ 
one may neglect $p_i$) \cite{Marchesini04}. This implies that 
one may assume {\em parton-flow} $\approx$~ {\em hadron-flow}.
Particles include both charged
particles measured by the tracker detector and clusters in the 
electromagnetic calorimeter.

More quantitatively, the observables $\Psi(r)$~ and  $\rho(r)$~ are defined 
as follows:

\beq{jshape}
\begin{array}{lll}
\Psi(r) & = &\int_0^r\,dr'\,\rho(r')\,,\\
\\
\rho(r) & = & 2\pi\, r \,\int_0^{E_T(jet)}\, dp_T\, \frac{p_T(r)}{E_T(jet)}\\
\\
& \times & \frac{d^2n(r,p_T)}{dr\,dp_T}\,,
\end{array}
\eeq

\npar where $r$~ is the distance to the jet axis in $(\eta,\phi)$~plane,
$p_T$ is the transverse momentum of a particle,
and $\frac{d^2n(r,p_T)}{dr\,dp_T}$~ is the particle number density 
over $r$~ and $p_T$.

In the present work we study also the dependence of jet structure on atomic
number of a target nucleus. In principle, there may exist an uncertainty 
with choice of the center of mass reference frame --- 
either $(pp)$, or $(pn)$~ or $(pA)$.
To avoid this problem we perform our analysis in the laboratory
reference frame.

Due to the above uncertainty with the choice of reference
frame we do not present the more traditional longitudinal and transversal
(with the respect of the jet axis) distributions of particles 
(``fragmentation function''). These observables are not invariant 
under longitudinal Lorentz boosts and usually are given 
in the beam-target center of mass system.

% ==============================================================

\section{HERA-B detector and data sample}
\label{sec:exper}

\begin{figure*}[ht]%          Fig. HERA-B detector
\centerline{%
\epsfig{figure=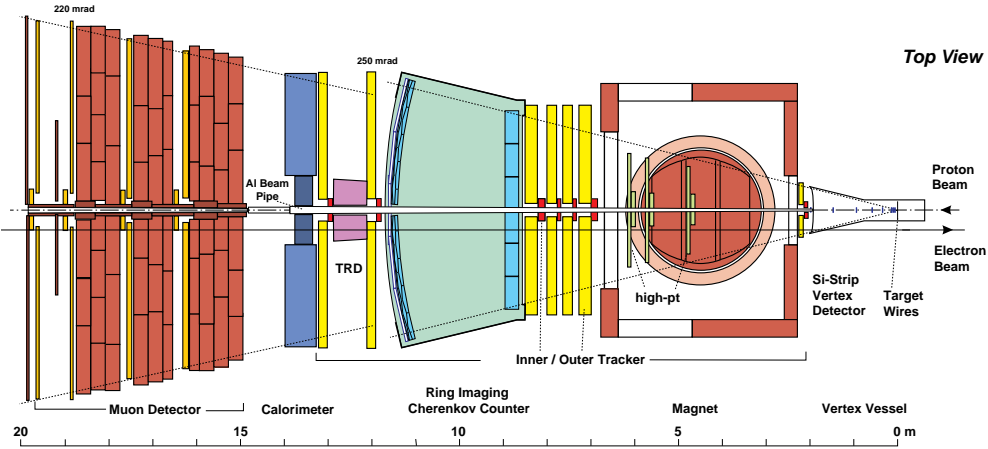,bbllx=-582pt,bblly=-5pt,%
bburx=59pt,bbury=288pt,clip=,height=6.0cm}}%
\caption{\label{herab} Schematic top view of the HERA-B detector}%
\end{figure*}

% - - - - - - - - - - - - - - - - - - - - - - - - - - - - - - -
\subsection{Detector}
\label{sec:det}

HERA-B is a fixed target experiment operated at the 920~GeV proton 
storage ring of HERA at DESY \cite{abt03}. 

A plan view of the HERA-B spectrometer is shown in  Fig.~\ref{herab}. 
The spectrometer dipole magnet provides a field integral of 2.13~T-m, with the 
main component perpendicular to the $x$-$z$~ plane. 
The apparatus (including particle identification
counters) has a forward acceptance of 15--220~mrad in the bending 
plane and 15--160~mrad in the non-bending plane.
The experiment uses a multi-wire fixed target which operates in the halo of 
the proton beam during HERA $e$-$p$~ collider operation. 
Up to eight different targets can be operated simultaneously, with their 
positions being adjusted dynamically in order to maintain a constant 
interaction rate between 1 and 40~MHz. 

The tracking system consists of a Vertex Detector System (VDS) \cite{VDS}
and a main tracker system. 
The VDS features 64 double-sided silicon microstrip detectors
arranged in eight stations along, and four quadrants around, 
the proton beam. 
The silicon strips have a readout pitch of approximately 50~$\mu$m.
Particle identification was performed by a Ring Imaging Cherenkov detector
(RICH) \cite{RICH}, an electromagnetic calorimeter (ECAL) \cite{ECAL} and
a muon detector (MUON) \cite{MUON}.

The main tracking system is separated into an Inner 
Tra\-cker (ITR) \cite{ITR} close to the proton beam-pipe and an Outer 
Tracker (OTR) \cite{OTR} farther out. 
The tracker system covers pseudorapidities within a range
of approximately $2\,\le\,\eta\,\le\,4.8$~ in the laboratory system.
Magnetic analysis for the OTR ensures the relative momentum resolution around
$\Delta p/p\,\approx\,5\cdot10^{-5}p\,\oplus\,1.6\cdot10^{-2}$.

The electromagnetic calorimeter ECAL is based on ``shashlyk'' 
sampling ca\-lo\-ri\-me\-ter tech\-nology, 
consisting of scintillator layers sandwiched between metal absorbers.
In the radially innermost section tungsten was used as an absorber,
and lead everywhere else. ECAL covers pseudorapidities within a range
of approximately $2\,\le\,\eta\,\le\,5.2$~ in laboratory system.
Calorimeter towers have a depth of about 20 radiation lengths with granularity 
approximately \\
$2.2\,\times\,2.2$~cm, $5.58\,\times\,5.58$~cm, 
and $11.15\,\times\,11.15$~cm for inner, middle and outer sections
respectively. Energy resolution $\Delta E/E$ is
$20.5\%/\sqrt{E}\,\oplus\,1.2\%$ for inner section and
approximately $11\%/\sqrt{E}\,\oplus\,1.0\%$ for middle and outer sections.

% - - - - - - - - - - - - - - - - - - - - - - - - - - - - - - -
\subsection{Data sample}
\label{sec:data}
 
For the analysis presented here, only data from VDS, 
OTR and the ECAL were used with carbon (C), aluminum (Al), 
and tungsten (W) wire targets.
The inner tra\-cker was not used due to its insufficient stability.

We used data collected in so called ``high $E_T$'' runs, 
i.e., runs in which about half of the events 
in each run were required to satisfy a calorimeter pre-trigger. 
The ``high $E_T$'' pre-trigger demands a transverse energy deposition 
at least in one of the ECAL towers above a predefined  threshold, 
$E_T\,\ge\,E_T(min)$. 
The remaining events in these runs were collected with a random
interaction trigger, in fact, minimum bias events. The data were recorded
at a moderate interaction rate, $1\,-\,3$~MHz, which corresponds
to $\approx\,0.1\,-\,0.3$~interactions per filled bunch crossing. 
Therefore only a small fraction of events contain more than one interaction.
Tungsten data were collected with pre-scaling by a factor $2$.

The results presented here are based on a sample of $\approx$~18~million events 
collected in the period from December 2002 to March 2003. 
There were three runs with a carbon target, two runs with a tungsten target,
and three runs with an aluminium target with  ECAL pre-trigger either
$E_T\,>\,3$~GeV or $E_T\,>\,2$~GeV.
Note that about 40\% of triggers were produced by ECAL clusters
matched with tracks/segments in tracking detector.

For the additional off-line event selection, we require that
the ECAL cluster with maximal transverse momentum (called in the following
the ``trigger cluster'') has $p^{max}_T\,\ge\,3.0$~GeV. 
The choice of the cluster with maximal $p_T$
unifies the selection of data and MC samples.
This ''trigger cluster'' does not necessarily coincide
with the ECAL cluster which produced the hardware pre-trigger. 
But the pre-trigger and the hardest ECAL clusters
differ in less than $0.1$\% of events.

Events were required to have a reconstructed primary vertex.
The track and cluster selection criteria are the following:

\begin{itemize}
\item Track must start in the VDS,
\item Track must be successfully fitted over the whole length (OTR and VDS),
\item ``Clones'' (nearby reconstructed tracks originating from the same
real physical track) are removed,
\item EM clusters must not match any track or track segment in the OTR,
\item The trigger conditions are tested before the check on track-cluster 
matching. If the trig\-ger clus\-ter mat\-ches a track or a track seg\-ment, 
the event is accepted, but the trigger cluster itself 
is removed from  further analysis.
\end{itemize}

After applying above selection criteria we have for our analysis approximately
1097000 carbon ev\-ents, 447000 tungsten events and 1002000 aluminium events.

We considered only the jet with the maximal transverse energy in the event.
Because our trigger is based on measuring the electromagnetic clusters
in the ECAL, it can distort properties of the selected jets,
preferring the jets enriched by gammas/$\pi^0$'s.
To avoid this problem, we accepted only ``away-side'' jets, i.e., jets
directed in the opposite direction with respect to the trigger cluster
in the plane perpendicular to the beam axis, demanding
$\vert\phi(jet)\,-\,\phi(trig)\vert\,\ge\,90^o$.
The Monte Carlo simulation shows that such a selection gives minimal
distortion of the jet transverse energy and its direction in comparison to
the parent parton. 

To minimize effects of the restricted acceptance 
we accept jets with axes in the narrow pseudo-ra\-pi\-di\-ty range 
$3.4\,\le\,\eta(jet)\,\le\,3.6$.

% ==============================================================
%\nextpage
\section{Monte Carlo simulation}
\label{sec:mc}

The simulation of the physical processes and the detector response 
in modern experiments is one of the crucial parts of data analysis.
For the simulation of jet events in HERA-B, the main problem
is that existing generators describing the hadron(nucleus)-nucleus 
collisions, are intended for the study of minimum bias physics and 
soft processes.  Therefore their use for simulation of hard 
hadron-nucleus collisions is extremely inefficient (if it is possible at all).
Due to these problems
a procedure for the description of hard parton scattering, taking
into account soft interactions in the nucleus was developed.

The bases of the developed MC generator are theoretical expectations as well
as experimental evidence \cite{Stewart90,Alverson94} that at moderate
ener\-gies $\sqrt{s}\approx 40$~GeV, the hard-scattering mechanism already
dominates $pp$~ collisions, whereas soft scattering is still the dominant
mechanism for the proton collisions with heavy nuclei. 
This interpretation is consistent
with the decrease of nuclear effects with increasing jet $E_T$.
All available data on the hadroproduction of jets are consistent with
a rather modest nuclear enhancement (the parameter $\alpha$~ in $A$-dependence
is close to unity, $\alpha\,<\,1.10$), diminishing with increasing jet $E_T$.

This fact allows us to discriminate hard and soft processes and
to exploit the existing MC models for hard parton scattering
to simulate jet production in proton-nucleus collisions. 

We use PYTHIA~6.2 \cite{Pythia} to produce hadronic systems in parton-parton
scattering, adding to the hard system soft particles produced in
the collision of the proton remnant with the nucleus. Note that
in PYTHIA, the partonic processes are simulated using LO matrix elements with
the inclusion of initial- and final-state parton showers. 

The program FRITIOF 7.02 \cite{Fritiof} from CERN library 
(including ARIADNE~4.02 as a part of FRI\-TI\-OF package) has been adapted 
to double precision to make it compatible with PYTHIA 6.2. 
Because the FRI\-TI\-OF package is heavily based on PYTHIA, 
all parameters for soft (semi-hard) processes in FRI\-TI\-OF
are the default parameters of PYTHIA~6.2 in the adapted version of FRI\-TI\-OF.

In such a combined PY\-THIA~6.2/FRI\-TI\-OF~7.02 package
the event simulation procedure consists of the following steps:

\begin{itemize}
\item simulation of the hard sub-process(es) (par\-ton-par\-ton scattering)
with PY\-THIA in $pN$~ collision
(a nucleon $N$ is being chosen randomly according to the nucleus content),
\item the independent fragmentation of scattered partons is performed,
\item resulting stable particles (hard system) are saved,
\item simulation of the soft collision is performed 
by the FRI\-TI\-OF generator as a $pA$ interaction 
with energy remaining after the hard scattering\footnote{We also tested 
the simulation of $\pi A$~ collisions for the remnant. 
The difference is insignificant.},
\item the transverse momenta of the whole system are balanced,
preserving the 4-momentum of the hard system,
\item the final fragmentation and decays are performed.
\end{itemize}

\vbox{%
The main PYTHIA inputs used for hard scattering are the following:
\begin{itemize}
\item Proton structure function ``CTEQ2L (best LO fit)'' from PDFLIB,
\item Intrinsic $k_T\,=\,1$~GeV of partons (default for PYTHIA~6.2), 
\item Independent fragmentation of final state partons from the hard process,
\item String fragmentation in soft interactions with the nucleus (in FRITIOF),
\item Lund symmetric fragmentation model both for independent and
string fragmentation,
\item QCD scale for parton-parton scattering as is the default for PYTHIA~6.2.
\end{itemize}
}

We tested that the default set of PYTHIA inputs gives a quite satisfactory 
description of our data as well as a good reproduction of the prompt photon
$p_T$ distribution measured by experiment E-706 \cite{E706} in 
$pBe$~ collisions at $\sqrt{s}\,=\,38.8$~GeV.

Note here that because we restricted our studies to the narrow
range $\Delta\eta\,=\,\pm 0.1$~ the variation of the PDF does not play 
any r\^ole. 
We also checked that the choice of the QCD scale for parton scattering 
is not significant in our range of transverse momenta and pseudo-rapidities.

The main effect comes from the intrinsic transverse momentum, $k_T$,
of partons. In paper \cite{E706} the authors suggest the value 
$k_T\,=\,1.3$~GeV to describe their results on prompt photon production.
But we did not find a significant difference between the values  
$k_T\,=\,1.3$~GeV
and the default PYTHIA~6.2 value, $k_T\,=\,1.0$~GeV for our conditions 
and use the default PYTHIA~6.2 value. 

To increase the simulation efficiency the parton sub-processes 
were simulated with the cut \\
\hbox{$p_T(hard)\,\ge\,3$~GeV,} where the $p_T(hard)$~ is the transverse 
momentum of the outgoing $\gamma$/parton. 
Such a cut increases the simulation efficiency by a few orders 
of magnitude  in comparison to the default cut in PYTHIA for hard scattering 
$p_T(hard)\,\ge\,1$~GeV.

The simulation of the detector response was carried out using the GEANT~3.21 
package with subsequent standard HERA-B reconstruction of the simulated 
events.

We applied the same selection criteria for tracks and EM clusters as for real 
data. In the MC sample we also removed from consideration (as in data) the 
so called ``hot modules'' in ECAL. ``Hot modules'' cover a negligibly 
small fraction of the total ECAL acceptance and, in fact, do not affect 
the presented results. 
In total, after selection (without trigger cut), the number of MC events 
is approximately equal to experimental statistics --- 1499000 events 
for the carbon 
target, 571400 events for the tungsten target and 1016600 events 
for the aluminium target. 

% ===========================================================
\section{Results}
\label{sec:result}

The measured, raw differential jet shape distribution $\rho(r)$~§was created 
by storing the entries in the bin $\Delta r$~ with weight $p_T/E_T(jet)$~ 
and further dividing this distribution by the bin width $\Delta r$. 

To correct the measured jet shape distributions for acceptance and
reconstruction distortions (systematic corrections) we used 
bin-by-bin corrections. Corrections for $\rho(r)$~ 
were calculated in each bin of jet transverse energy
$E_T$~ separately and can be written as:

\beq{corrs}
\begin{array}{lll}
\rho(r;E_T) & = & R(r;E_T)\,\times\,\rho_{m}(r;E_T)\\
R(r;E_T)    & = & \rho_{g}(r;E_T)/\rho_{r}(r;E_T)
\end{array}
\eeq

\npar where, $\rho$~ and $\rho_{m}$~ are corrected and measured
distributions and $\rho_{r}$~ and $\rho_{g}$~ are simulated distributions
at the reconstructed and generator levels, respectively.

We tested the quality of the track information to check for possible
systematic biases in the track momentum measurements.
For this purpose we reconstructed known resonances $K^0$, $\rho^0$, $K^{*0}$, 
$\phi^0$, and $\Lambda^0$. Results for reconstructed masses
are in good agreement with the known values of the Particle Data Group 
\cite{PDG04}. 
Thus we don't expect any significant systematic bias in measured track momenta.

We checked, that after applying the trigger cut $p^{max}_T\,>\,3$~GeV, 
the simulated distributions for jets and particles agree well
with data except for the multiplicity distributions for low-$p_T$~ particles, 
where simulated events have lower average multiplicity. This discrepancy
is a known problem of the FRITIOF generator.
But this discrepancy does not distort jet shape distribution
in the selected jets and does not play a significant r\^ole in bin-by-bin
correction procedure.

The trigger cut, $p^{max}_T\,>\,3$~GeV, leaves about $0.5$\% of the total MC 
sample. Using the simulated sample, we checked that the event triggering 
does not distort the jet shape for ``away-side'' jets. 
We checked also that triggered and non-triggered MC samples both give good
description of the data jet shape within respective statistical errors. 
This fact allows us to use the non-trig\-ge\-red MC sample to find systematic 
cor\-rec\-tions $R(r;E_T)$, thus reducing statistical systematic errors
by more than an order of magnitude.

We performed measurements of the jet shape in the range of jet transverse
energies $E_T\,>\,5$~GeV, 
to minimize the in\-flu\-ence of the cut $p_T(hard)\,>\,3$~GeV 
which we applied when simulating par\-ton-par\-ton scat\-tering. 
Due to the presence of the Gaus\-sian-dist\-ributed intrinsic transverse 
mo\-men\-tum with width $k_T\,=\,1$~GeV, the missing range of parton transverse
momenta, $p_T\,<\,3$~GeV, gives a negligible contribution to the jet yield
for $E_T\,>\,5$~GeV. In this case one can expect that MC based systematic
corrections give reliable results.

We could select jets up to transverse energy $E_T\,\le\,14$~ GeV,
however, for $E_T\,>\,10$~GeV, the statistics is small due to the steeply 
falling jet $E_T$~ spectrum. Therefore,
we divide $E_T(jet)$ range $5\,-\,10$~GeV into five equal bins  
and chose one bin for $E_T\,>\,10$~GeV. We apply bin-by-bin corrections
(\ref{corrs}) to measured differential jet shape $\rho(r)$ 
in each $E_T(jet)$~ bin.

Our results for differential and integral jet shapes
are presented in Tables \ref{car:1}, \ref{car:2} for the carbon
target, in Tables \ref{alu:1}, \ref{alu:2} for the aluminium target
and in Tables \ref{tun:1}, \ref{tun:2} for the tungsten target. 
We take as a systematic error the uncertainty on the correction factors 
defined by the finite Monte Carlo statistics. Our studies show that
MC uncertainties dominate the other possible sources of systematic errors. 
Errors given in the tables are the quadratic sum of statistical
and systematic errors. Both types of errors are, approximately, equal
each to other. 
The corrected $\Psi(r;E_T)$~ distribution was obtained by integration of
the corrected $\rho(r;E_T)$~ distribution in each $E_T(jet)$~ bin according to:

\bmath
\Psi(r_n;E_T)\ = \ \sum_{i=1}^{n}\,\rho(r_i;E_T)\,\Delta r\,,
\emath

\npar where $i$~ denotes the bin in a histogram.

Results of measurements of $\rho(r)$~ and $\Psi(r)$~ 
are presented in Tables \ref{car:1}--\ref{tun:2}.

% ===========================================================
%\nextpage
\begin{table*}
\caption{Differential and integral jet shapes for carbon for $5<E_T<8$~GeV.}
\label{car:1}       % Give a unique label
% For LaTeX tables use
\begin{tabular}{|r|rr|rr|rr|}
\hline
 && $E_T=5-6$~GeV  && $E_T=6-7$~GeV && $E_T=7-8$~GeV \\
\hline
r & $\rho$ & $\Psi$   & $\rho$ & $\Psi$ & $\rho$ & $\Psi$ \\
\hline
\hline
0.05 & 0.550 $\pm$ 0.040 & 0.055 $\pm$ 0.004 & 0.550 $\pm$ 0.050 & 0.055 $\pm$ 0.005 & 0.420 $\pm$ 0.050 & 0.042 $\pm$ 0.005 \\
0.15 & 1.250 $\pm$ 0.050 & 0.180 $\pm$ 0.006 & 1.210 $\pm$ 0.060 & 0.176 $\pm$ 0.008 & 1.050 $\pm$ 0.080 & 0.147 $\pm$ 0.009 \\
0.25 & 1.510 $\pm$ 0.050 & 0.331 $\pm$ 0.008 & 1.390 $\pm$ 0.060 & 0.320 $\pm$ 0.010 & 1.500 $\pm$ 0.100 & 0.295 $\pm$ 0.014 \\
0.35 & 1.440 $\pm$ 0.040 & 0.476 $\pm$ 0.009 & 1.410 $\pm$ 0.050 & 0.460 $\pm$ 0.010 & 1.430 $\pm$ 0.080 & 0.438 $\pm$ 0.016 \\
0.45 & 1.370 $\pm$ 0.040 & 0.613 $\pm$ 0.009 & 1.410 $\pm$ 0.050 & 0.600 $\pm$ 0.010 & 1.420 $\pm$ 0.070 & 0.580 $\pm$ 0.020 \\
0.55 & 1.200 $\pm$ 0.030 & 0.730 $\pm$ 0.010 & 1.120 $\pm$ 0.040 & 0.710 $\pm$ 0.010 & 1.120 $\pm$ 0.060 & 0.690 $\pm$ 0.020 \\
0.65 & 0.880 $\pm$ 0.020 & 0.820 $\pm$ 0.010 & 0.890 $\pm$ 0.030 & 0.800 $\pm$ 0.010 & 0.900 $\pm$ 0.040 & 0.780 $\pm$ 0.020 \\
0.75 & 0.630 $\pm$ 0.020 & 0.880 $\pm$ 0.010 & 0.680 $\pm$ 0.020 & 0.866 $\pm$ 0.014 & 0.750 $\pm$ 0.040 & 0.860 $\pm$ 0.020 \\
0.85 & 0.460 $\pm$ 0.010 & 0.930 $\pm$ 0.010 & 0.490 $\pm$ 0.020 & 0.915 $\pm$ 0.014 & 0.470 $\pm$ 0.020 & 0.900 $\pm$ 0.020 \\
0.95 & 0.287 $\pm$ 0.009 & 0.960 $\pm$ 0.010 & 0.358 $\pm$ 0.014 & 0.951 $\pm$ 0.014 & 0.380 $\pm$ 0.020 & 0.940 $\pm$ 0.020 \\
1.05 & 0.177 $\pm$ 0.006 & 0.980 $\pm$ 0.010 & 0.220 $\pm$ 0.010 & 0.973 $\pm$ 0.014 & 0.242 $\pm$ 0.015 & 0.970 $\pm$ 0.020 \\
1.15 & 0.112 $\pm$ 0.005 & 0.990 $\pm$ 0.010 & 0.122 $\pm$ 0.006 & 0.985 $\pm$ 0.014 & 0.150 $\pm$ 0.010 & 0.980 $\pm$ 0.020 \\
1.25 & 0.057 $\pm$ 0.003 & 0.990 $\pm$ 0.010 & 0.076 $\pm$ 0.004 & 0.992 $\pm$ 0.014 & 0.097 $\pm$ 0.008 & 0.990 $\pm$ 0.020 \\
1.35 & 0.032 $\pm$ 0.002 & 1.000 $\pm$ 0.010 & 0.039 $\pm$ 0.003 & 0.996 $\pm$ 0.014 & 0.049 $\pm$ 0.005 & 1.000 $\pm$ 0.020 \\
1.45 & 0.017 $\pm$ 0.001 &                   & 0.019 $\pm$ 0.002 & 0.998 $\pm$ 0.014 & 0.024 $\pm$ 0.003 &                   \\
1.55 & 0.008 $\pm$ 0.001 &                   & 0.011 $\pm$ 0.001 & 0.999 $\pm$ 0.014 & 0.011 $\pm$ 0.002 &                   \\
1.65 & 0.003 $\pm$ 0.001 &                   & 0.005 $\pm$ 0.001 & 1.000 $\pm$ 0.014 & 0.006 $\pm$ 0.001 &                   \\
1.75 & 0.001 $\pm$ 0.001 &                   & 0.001 $\pm$ 0.001 &                   & 0.003 $\pm$ 0.001 &                   \\
1.85 &                   &                   & 0.001 $\pm$ 0.001 &                   & 0.001 $\pm$ 0.001 &                   \\

\hline
\end{tabular}
\end{table*}

% ===========================================================
%\nextpage
\begin{table*}
\caption{Differential and integral jet shapes for carbon for $E_T>8$~GeV.}
\label{car:2}       % Give a unique label
% For LaTeX tables use
\begin{tabular}{|r|rr|rr|rr|}
\hline
 && $E_T=8-9$~GeV  && $E_T=9-10$~GeV && $E_T>10$~GeV \\
\hline
r & $\rho$ & $\Psi$   & $\rho$ & $\Psi$ & $\rho$ & $\Psi$ \\
\hline
\hline
0.05 & 0.280 $\pm$ 0.040 & 0.028 $\pm$ 0.004 & 0.280 $\pm$ 0.050 & 0.027 $\pm$ 0.005 & 0.150 $\pm$ 0.060 & 0.015 $\pm$ 0.006 \\
0.15 & 0.900 $\pm$ 0.100 & 0.121 $\pm$ 0.010 & 0.700 $\pm$ 0.100 & 0.100 $\pm$ 0.010 & 1.000 $\pm$ 0.300 & 0.120 $\pm$ 0.030 \\
0.25 & 1.350 $\pm$ 0.140 & 0.256 $\pm$ 0.020 & 1.140 $\pm$ 0.160 & 0.220 $\pm$ 0.020 & 1.600 $\pm$ 0.300 & 0.280 $\pm$ 0.040 \\
0.35 & 1.300 $\pm$ 0.100 & 0.383 $\pm$ 0.020 & 1.600 $\pm$ 0.200 & 0.370 $\pm$ 0.030 & 1.400 $\pm$ 0.300 & 0.420 $\pm$ 0.050 \\
0.45 & 1.400 $\pm$ 0.100 & 0.529 $\pm$ 0.020 & 1.600 $\pm$ 0.200 & 0.530 $\pm$ 0.040 & 1.400 $\pm$ 0.200 & 0.560 $\pm$ 0.050 \\
0.55 & 1.220 $\pm$ 0.090 & 0.651 $\pm$ 0.030 & 1.200 $\pm$ 0.100 & 0.650 $\pm$ 0.040 & 1.130 $\pm$ 0.150 & 0.670 $\pm$ 0.060 \\
0.65 & 1.040 $\pm$ 0.070 & 0.755 $\pm$ 0.030 & 0.960 $\pm$ 0.090 & 0.750 $\pm$ 0.040 & 0.670 $\pm$ 0.090 & 0.740 $\pm$ 0.060 \\
0.75 & 0.790 $\pm$ 0.060 & 0.835 $\pm$ 0.030 & 0.650 $\pm$ 0.060 & 0.810 $\pm$ 0.040 & 0.640 $\pm$ 0.090 & 0.800 $\pm$ 0.060 \\
0.85 & 0.550 $\pm$ 0.040 & 0.889 $\pm$ 0.030 & 0.620 $\pm$ 0.060 & 0.880 $\pm$ 0.040 & 0.560 $\pm$ 0.080 & 0.860 $\pm$ 0.060 \\
0.95 & 0.430 $\pm$ 0.030 & 0.933 $\pm$ 0.030 & 0.380 $\pm$ 0.040 & 0.910 $\pm$ 0.040 & 0.470 $\pm$ 0.080 & 0.910 $\pm$ 0.060 \\
1.05 & 0.260 $\pm$ 0.020 & 0.959 $\pm$ 0.030 & 0.340 $\pm$ 0.040 & 0.950 $\pm$ 0.040 & 0.360 $\pm$ 0.060 & 0.940 $\pm$ 0.060 \\
1.15 & 0.200 $\pm$ 0.020 & 0.979 $\pm$ 0.030 & 0.220 $\pm$ 0.030 & 0.970 $\pm$ 0.040 & 0.130 $\pm$ 0.030 & 0.960 $\pm$ 0.060 \\
1.25 & 0.100 $\pm$ 0.010 & 0.989 $\pm$ 0.030 & 0.130 $\pm$ 0.020 & 0.980 $\pm$ 0.040 & 0.190 $\pm$ 0.040 & 0.970 $\pm$ 0.060 \\
1.35 & 0.050 $\pm$ 0.007 & 0.994 $\pm$ 0.030 & 0.084 $\pm$ 0.015 & 0.990 $\pm$ 0.040 & 0.130 $\pm$ 0.030 & 0.990 $\pm$ 0.060 \\
1.45 & 0.023 $\pm$ 0.004 & 0.996 $\pm$ 0.030 & 0.050 $\pm$ 0.010 & 1.000 $\pm$ 0.040 & 0.050 $\pm$ 0.020 & 0.990 $\pm$ 0.060 \\
1.55 & 0.015 $\pm$ 0.003 & 0.998 $\pm$ 0.030 & 0.028 $\pm$ 0.009 &                   & 0.040 $\pm$ 0.010 & 1.000 $\pm$ 0.060 \\
1.65 & 0.015 $\pm$ 0.006 & 0.999 $\pm$ 0.030 & 0.015 $\pm$ 0.006 &                   & 0.040 $\pm$ 0.020 &                   \\
1.75 & 0.002 $\pm$ 0.001 & 1.000 $\pm$ 0.030 & 0.005 $\pm$ 0.002 &                   & 0.007 $\pm$ 0.004 &                   \\
1.85 & 0.002 $\pm$ 0.002 &                   & 0.004 $\pm$ 0.003 &                   & 0.008 $\pm$ 0.005 &                   \\
1.95 &                   &                   &                   &                   & 0.003 $\pm$ 0.003 &                   \\

\hline
\end{tabular}
\end{table*}

% ===========================================================
%\nextpage
\begin{table*}
\caption{Differential and integral jet shapes for aluminium $5<E_T<8$~GeV.}
\label{alu:1}       % Give a unique label
% For LaTeX tables use
\begin{tabular}{|r|rr|rr|rr|}
\hline
 && $E_T=5-6$~GeV  && $E_T=6-7$~GeV && $E_T=7-8$~GeV \\
\hline
r & $\rho$ & $\Psi$   & $\rho$ & $\Psi$ & $\rho$ & $\Psi$ \\
\hline
\hline
0.05 & 0.660 $\pm$ 0.050 & 0.066 $\pm$ 0.005 & 0.580 $\pm$ 0.060 & 0.058 $\pm$ 0.006 & 0.430 $\pm$ 0.050 & 0.043 $\pm$ 0.005 \\
0.15 & 1.230 $\pm$ 0.050 & 0.189 $\pm$ 0.007 & 1.220 $\pm$ 0.070 & 0.179 $\pm$ 0.009 & 1.160 $\pm$ 0.100 & 0.160 $\pm$ 0.010 \\
0.25 & 1.480 $\pm$ 0.050 & 0.337 $\pm$ 0.009 & 1.510 $\pm$ 0.080 & 0.330 $\pm$ 0.010 & 1.400 $\pm$ 0.100 & 0.303 $\pm$ 0.016 \\
0.35 & 1.450 $\pm$ 0.040 & 0.480 $\pm$ 0.010 & 1.420 $\pm$ 0.060 & 0.470 $\pm$ 0.010 & 1.480 $\pm$ 0.090 & 0.450 $\pm$ 0.020 \\
0.45 & 1.380 $\pm$ 0.040 & 0.620 $\pm$ 0.010 & 1.340 $\pm$ 0.050 & 0.610 $\pm$ 0.010 & 1.250 $\pm$ 0.070 & 0.580 $\pm$ 0.020 \\
0.55 & 1.120 $\pm$ 0.030 & 0.730 $\pm$ 0.010 & 1.150 $\pm$ 0.040 & 0.722 $\pm$ 0.015 & 1.060 $\pm$ 0.060 & 0.680 $\pm$ 0.020 \\
0.65 & 0.870 $\pm$ 0.020 & 0.820 $\pm$ 0.010 & 0.850 $\pm$ 0.030 & 0.806 $\pm$ 0.015 & 0.950 $\pm$ 0.050 & 0.780 $\pm$ 0.020 \\
0.75 & 0.660 $\pm$ 0.020 & 0.880 $\pm$ 0.010 & 0.670 $\pm$ 0.020 & 0.874 $\pm$ 0.016 & 0.750 $\pm$ 0.040 & 0.850 $\pm$ 0.020 \\
0.85 & 0.450 $\pm$ 0.014 & 0.930 $\pm$ 0.010 & 0.480 $\pm$ 0.020 & 0.922 $\pm$ 0.016 & 0.560 $\pm$ 0.030 & 0.910 $\pm$ 0.020 \\
0.95 & 0.290 $\pm$ 0.010 & 0.960 $\pm$ 0.010 & 0.310 $\pm$ 0.010 & 0.953 $\pm$ 0.016 & 0.350 $\pm$ 0.020 & 0.940 $\pm$ 0.020 \\
1.05 & 0.194 $\pm$ 0.008 & 0.980 $\pm$ 0.010 & 0.200 $\pm$ 0.010 & 0.973 $\pm$ 0.016 & 0.247 $\pm$ 0.016 & 0.970 $\pm$ 0.020 \\
1.15 & 0.111 $\pm$ 0.005 & 0.990 $\pm$ 0.010 & 0.123 $\pm$ 0.007 & 0.985 $\pm$ 0.016 & 0.150 $\pm$ 0.010 & 0.980 $\pm$ 0.020 \\
1.25 & 0.063 $\pm$ 0.003 & 0.990 $\pm$ 0.010 & 0.074 $\pm$ 0.005 & 0.993 $\pm$ 0.016 & 0.086 $\pm$ 0.007 & 0.990 $\pm$ 0.020 \\
1.35 & 0.030 $\pm$ 0.002 & 1.000 $\pm$ 0.010 & 0.038 $\pm$ 0.003 & 0.996 $\pm$ 0.016 & 0.042 $\pm$ 0.005 & 1.000 $\pm$ 0.020 \\
1.45 & 0.016 $\pm$ 0.001 &                   & 0.022 $\pm$ 0.002 & 0.999 $\pm$ 0.016 & 0.024 $\pm$ 0.003 &                   \\
1.55 & 0.008 $\pm$ 0.001 &                   & 0.009 $\pm$ 0.001 & 0.999 $\pm$ 0.016 & 0.009 $\pm$ 0.002 &                   \\
1.65 &                   &		     & 0.004 $\pm$ 0.001 & 1.000 $\pm$ 0.016 & 0.009 $\pm$ 0.002 &                   \\
1.75 &                   &		     & 0.001 $\pm$ 0.001 &		     & 0.004 $\pm$ 0.001 &                   \\
1.85 &                   &                   &                   &                   & 0.002 $\pm$ 0.001 &                   \\
1.95 &                   &                   &                   &                   & 0.001 $\pm$ 0.001 &                   \\

\hline
\end{tabular}
\end{table*}

% ===========================================================
%\nextpage
\begin{table*}
\caption{Differential and integral jet shapes for aluminium $E_T>8$~GeV.}
\label{alu:2}       % Give a unique label
% For LaTeX tables use
\begin{tabular}{|r|rr|rr|rr|}
\hline
 && $E_T=8-9$~GeV  && $E_T=9-10$~GeV && $E_T>10$~GeV \\
\hline
r & $\rho$ & $\Psi$   & $\rho$ & $\Psi$ & $\rho$ & $\Psi$ \\
\hline
\hline
0.05 & 0.340 $\pm$ 0.050 & 0.034 $\pm$ 0.005 & 0.300 $\pm$ 0.060 & 0.030 $\pm$ 0.006 & 0.170 $\pm$ 0.060 & 0.017 $\pm$ 0.006 \\
0.15 & 1.000 $\pm$ 0.100 & 0.130 $\pm$ 0.010 & 0.600 $\pm$ 0.100 & 0.090 $\pm$ 0.010 & 0.900 $\pm$ 0.200 & 0.110 $\pm$ 0.020 \\
0.25 & 1.440 $\pm$ 0.150 & 0.280 $\pm$ 0.020 & 1.300 $\pm$ 0.200 & 0.220 $\pm$ 0.020 & 1.200 $\pm$ 0.200 & 0.230 $\pm$ 0.030 \\
0.35 & 1.300 $\pm$ 0.100 & 0.400 $\pm$ 0.020 & 1.300 $\pm$ 0.200 & 0.360 $\pm$ 0.030 & 1.600 $\pm$ 0.300 & 0.390 $\pm$ 0.040 \\
0.45 & 1.300 $\pm$ 0.100 & 0.530 $\pm$ 0.030 & 1.300 $\pm$ 0.150 & 0.490 $\pm$ 0.030 & 1.000 $\pm$ 0.200 & 0.490 $\pm$ 0.040 \\
0.55 & 1.300 $\pm$ 0.100 & 0.660 $\pm$ 0.030 & 1.300 $\pm$ 0.100 & 0.610 $\pm$ 0.040 & 1.200 $\pm$ 0.200 & 0.610 $\pm$ 0.050 \\
0.65 & 1.030 $\pm$ 0.080 & 0.760 $\pm$ 0.030 & 1.000 $\pm$ 0.100 & 0.710 $\pm$ 0.040 & 0.900 $\pm$ 0.100 & 0.700 $\pm$ 0.050 \\
0.75 & 0.740 $\pm$ 0.050 & 0.830 $\pm$ 0.030 & 0.900 $\pm$ 0.100 & 0.800 $\pm$ 0.040 & 0.550 $\pm$ 0.080 & 0.750 $\pm$ 0.050 \\
0.85 & 0.500 $\pm$ 0.040 & 0.880 $\pm$ 0.030 & 0.600 $\pm$ 0.060 & 0.860 $\pm$ 0.040 & 0.700 $\pm$ 0.200 & 0.820 $\pm$ 0.060 \\
0.95 & 0.480 $\pm$ 0.040 & 0.930 $\pm$ 0.030 & 0.540 $\pm$ 0.060 & 0.920 $\pm$ 0.040 & 0.600 $\pm$ 0.100 & 0.880 $\pm$ 0.060 \\
1.05 & 0.270 $\pm$ 0.020 & 0.960 $\pm$ 0.030 & 0.330 $\pm$ 0.040 & 0.950 $\pm$ 0.040 & 0.400 $\pm$ 0.100 & 0.920 $\pm$ 0.060 \\
1.15 & 0.152 $\pm$ 0.015 & 0.970 $\pm$ 0.030 & 0.180 $\pm$ 0.030 & 0.970 $\pm$ 0.040 & 0.400 $\pm$ 0.100 & 0.960 $\pm$ 0.060 \\
1.25 & 0.121 $\pm$ 0.014 & 0.990 $\pm$ 0.030 & 0.160 $\pm$ 0.030 & 0.980 $\pm$ 0.040 & 0.180 $\pm$ 0.030 & 0.980 $\pm$ 0.060 \\
1.35 & 0.070 $\pm$ 0.010 & 0.990 $\pm$ 0.030 & 0.082 $\pm$ 0.016 & 0.990 $\pm$ 0.040 & 0.100 $\pm$ 0.020 & 0.990 $\pm$ 0.060 \\
1.45 & 0.037 $\pm$ 0.007 & 1.000 $\pm$ 0.030 & 0.048 $\pm$ 0.010 & 1.000 $\pm$ 0.040 & 0.060 $\pm$ 0.020 & 1.000 $\pm$ 0.060 \\
1.55 & 0.024 $\pm$ 0.006 &                   & 0.025 $\pm$ 0.007 &                   & 0.030 $\pm$ 0.010 &                   \\
1.65 & 0.008 $\pm$ 0.002 &                   & 0.010 $\pm$ 0.004 &                   & 0.008 $\pm$ 0.003 &                   \\
1.75 & 0.003 $\pm$ 0.001 &                   & 0.001 $\pm$ 0.001 &		     & 0.007 $\pm$ 0.003 &                   \\
1.85 & 0.005 $\pm$ 0.003 &                   &			 &		     &			 &		     \\
1.95 & 0.001 $\pm$ 0.001 &                   &                   &                   &                   &                   \\

\hline
\end{tabular}
\end{table*}

% ===========================================================
%\nextpage
\begin{table*}
\caption{Differential and integral jet shapes for tungsten $5<E_T<8$~GeV.}
\label{tun:1}       % Give a unique label
% For LaTeX tables use
\begin{tabular}{|r|rr|rr|rr|}
\hline
 && $E_T=5-6$~GeV  && $E_T=6-7$~GeV && $E_T=7-8$~GeV \\
\hline
r & $\rho$ & $\Psi$   & $\rho$ & $\Psi$ & $\rho$ & $\Psi$ \\
\hline
\hline
0.05 & 0.490 $\pm$ 0.050 & 0.049 $\pm$ 0.005 & 0.390 $\pm$ 0.040 & 0.039 $\pm$ 0.004 & 0.280 $\pm$ 0.040 & 0.028 $\pm$ 0.004 \\
0.15 & 1.040 $\pm$ 0.060 & 0.154 $\pm$ 0.008 & 1.030 $\pm$ 0.070 & 0.142 $\pm$ 0.009 & 0.790 $\pm$ 0.070 & 0.107 $\pm$ 0.008 \\
0.25 & 1.390 $\pm$ 0.060 & 0.290 $\pm$ 0.010 & 1.380 $\pm$ 0.080 & 0.280 $\pm$ 0.010 & 1.200 $\pm$ 0.100 & 0.230 $\pm$ 0.010 \\
0.35 & 1.580 $\pm$ 0.060 & 0.450 $\pm$ 0.010 & 1.460 $\pm$ 0.070 & 0.430 $\pm$ 0.010 & 1.460 $\pm$ 0.090 & 0.374 $\pm$ 0.016 \\
0.45 & 1.420 $\pm$ 0.050 & 0.590 $\pm$ 0.010 & 1.430 $\pm$ 0.060 & 0.569 $\pm$ 0.015 & 1.570 $\pm$ 0.090 & 0.530 $\pm$ 0.020 \\
0.55 & 1.280 $\pm$ 0.040 & 0.720 $\pm$ 0.010 & 1.220 $\pm$ 0.050 & 0.691 $\pm$ 0.016 & 1.320 $\pm$ 0.070 & 0.660 $\pm$ 0.020 \\
0.65 & 1.000 $\pm$ 0.040 & 0.820 $\pm$ 0.010 & 1.080 $\pm$ 0.040 & 0.800 $\pm$ 0.020 & 1.180 $\pm$ 0.060 & 0.780 $\pm$ 0.020 \\
0.75 & 0.730 $\pm$ 0.030 & 0.894 $\pm$ 0.015 & 0.760 $\pm$ 0.030 & 0.880 $\pm$ 0.020 & 0.780 $\pm$ 0.040 & 0.860 $\pm$ 0.020 \\
0.85 & 0.460 $\pm$ 0.020 & 0.940 $\pm$ 0.015 & 0.510 $\pm$ 0.020 & 0.930 $\pm$ 0.020 & 0.550 $\pm$ 0.030 & 0.910 $\pm$ 0.020 \\
0.95 & 0.280 $\pm$ 0.010 & 0.968 $\pm$ 0.015 & 0.340 $\pm$ 0.020 & 0.960 $\pm$ 0.020 & 0.370 $\pm$ 0.020 & 0.950 $\pm$ 0.020 \\
1.05 & 0.154 $\pm$ 0.008 & 0.983 $\pm$ 0.015 & 0.180 $\pm$ 0.010 & 0.980 $\pm$ 0.020 & 0.223 $\pm$ 0.016 & 0.970 $\pm$ 0.020 \\
1.15 & 0.088 $\pm$ 0.005 & 0.992 $\pm$ 0.015 & 0.114 $\pm$ 0.008 & 0.990 $\pm$ 0.020 & 0.140 $\pm$ 0.010 & 0.990 $\pm$ 0.020 \\
1.25 & 0.043 $\pm$ 0.003 & 0.996 $\pm$ 0.015 & 0.046 $\pm$ 0.004 & 1.000 $\pm$ 0.020 & 0.067 $\pm$ 0.007 & 0.990 $\pm$ 0.020 \\
1.35 & 0.022 $\pm$ 0.002 & 0.998 $\pm$ 0.015 & 0.031 $\pm$ 0.003 &                   & 0.037 $\pm$ 0.005 & 1.000 $\pm$ 0.020 \\
1.45 & 0.009 $\pm$ 0.001 & 0.999 $\pm$ 0.015 & 0.014 $\pm$ 0.002 &                   & 0.010 $\pm$ 0.002 &                   \\
1.55 & 0.004 $\pm$ 0.001 & 1.000 $\pm$ 0.015 & 0.005 $\pm$ 0.001 &                   & 0.008 $\pm$ 0.002 &                   \\
1.65 & 0.002 $\pm$ 0.001 &                   & 0.002 $\pm$ 0.001 &                   & 0.002 $\pm$ 0.001 &                   \\
1.75 & 0.001 $\pm$ 0.001 &                   & 0.001 $\pm$ 0.001 &                   & 0.002 $\pm$ 0.001 &                   \\
1.85 &                   &                   &                   &                   & 0.001 $\pm$ 0.001 &                   \\

\hline
\end{tabular}
\end{table*}

% ===========================================================
%\nextpage
\begin{table*}
\caption{Differential and integral jet shapes for tungsten $E_T>8$~GeV.}
\label{tun:2}       % Give a unique label
% For LaTeX tables use
\begin{tabular}{|r|rr|rr|rr|}
\hline
 && $E_T=8-9$~GeV  && $E_T=9-10$~GeV && $E_T>10$~GeV \\
\hline
r & $\rho$ & $\Psi$   & $\rho$ & $\Psi$ & $\rho$ & $\Psi$ \\
\hline
\hline
0.05 & 0.180 $\pm$ 0.030 & 0.017 $\pm$ 0.003 & 0.120 $\pm$ 0.020 & 0.012 $\pm$ 0.002 & 0.210 $\pm$ 0.160 & 0.021 $\pm$ 0.016 \\
0.15 & 0.640 $\pm$ 0.080 & 0.081 $\pm$ 0.008 & 0.560 $\pm$ 0.090 & 0.068 $\pm$ 0.009 & 0.540 $\pm$ 0.160 & 0.070 $\pm$ 0.020 \\
0.25 & 1.200 $\pm$ 0.100 & 0.198 $\pm$ 0.015 & 1.500 $\pm$ 0.200 & 0.220 $\pm$ 0.030 & 1.700 $\pm$ 0.500 & 0.250 $\pm$ 0.050 \\
0.35 & 1.650 $\pm$ 0.150 & 0.360 $\pm$ 0.020 & 1.400 $\pm$ 0.200 & 0.360 $\pm$ 0.030 & 1.500 $\pm$ 0.200 & 0.400 $\pm$ 0.060 \\
0.45 & 1.610 $\pm$ 0.140 & 0.520 $\pm$ 0.030 & 1.330 $\pm$ 0.150 & 0.490 $\pm$ 0.040 & 1.300 $\pm$ 0.200 & 0.530 $\pm$ 0.060 \\
0.55 & 1.260 $\pm$ 0.090 & 0.650 $\pm$ 0.030 & 1.400 $\pm$ 0.140 & 0.630 $\pm$ 0.040 & 1.200 $\pm$ 0.140 & 0.650 $\pm$ 0.060 \\
0.65 & 1.120 $\pm$ 0.080 & 0.760 $\pm$ 0.030 & 1.020 $\pm$ 0.100 & 0.730 $\pm$ 0.040 & 0.900 $\pm$ 0.100 & 0.740 $\pm$ 0.060 \\
0.75 & 0.740 $\pm$ 0.050 & 0.840 $\pm$ 0.030 & 0.930 $\pm$ 0.090 & 0.830 $\pm$ 0.040 & 0.800 $\pm$ 0.100 & 0.820 $\pm$ 0.060 \\
0.85 & 0.620 $\pm$ 0.050 & 0.900 $\pm$ 0.030 & 0.640 $\pm$ 0.070 & 0.890 $\pm$ 0.040 & 0.460 $\pm$ 0.060 & 0.870 $\pm$ 0.060 \\
0.95 & 0.420 $\pm$ 0.040 & 0.940 $\pm$ 0.030 & 0.440 $\pm$ 0.040 & 0.940 $\pm$ 0.040 & 0.430 $\pm$ 0.070 & 0.910 $\pm$ 0.060 \\
1.05 & 0.280 $\pm$ 0.030 & 0.970 $\pm$ 0.030 & 0.240 $\pm$ 0.030 & 0.960 $\pm$ 0.040 & 0.300 $\pm$ 0.040 & 0.940 $\pm$ 0.060 \\
1.15 & 0.152 $\pm$ 0.016 & 0.980 $\pm$ 0.030 & 0.160 $\pm$ 0.020 & 0.970 $\pm$ 0.040 & 0.280 $\pm$ 0.050 & 0.970 $\pm$ 0.060 \\
1.25 & 0.080 $\pm$ 0.010 & 0.990 $\pm$ 0.030 & 0.140 $\pm$ 0.020 & 0.990 $\pm$ 0.040 & 0.130 $\pm$ 0.030 & 0.980 $\pm$ 0.060 \\
1.35 & 0.045 $\pm$ 0.007 & 1.000 $\pm$ 0.030 & 0.060 $\pm$ 0.010 & 0.990 $\pm$ 0.040 & 0.100 $\pm$ 0.020 & 0.990 $\pm$ 0.060 \\
1.45 & 0.032 $\pm$ 0.006 &                   & 0.021 $\pm$ 0.005 & 1.000 $\pm$ 0.040 & 0.030 $\pm$ 0.010 & 0.990 $\pm$ 0.060 \\
1.55 & 0.014 $\pm$ 0.003 &                   & 0.024 $\pm$ 0.008 &                   & 0.024 $\pm$ 0.009 & 1.000 $\pm$ 0.060 \\
1.65 & 0.005 $\pm$ 0.002 &                   & 0.007 $\pm$ 0.003 &                   & 0.021 $\pm$ 0.009 &                   \\
1.75 & 0.002 $\pm$ 0.001 &                   & 0.004 $\pm$ 0.002 &                   & 0.003 $\pm$ 0.002 &                   \\

\hline
\end{tabular}
\end{table*}

%                       <<<<==========  Tables  ==============

\begin{figure*}[htb]
%\centerline{\psfig{figure=rhos.eps,height=9.0cm,width=16.0cm}}%
\begin{center}
\vbox{%
\hbox{%
\epsfig{figure=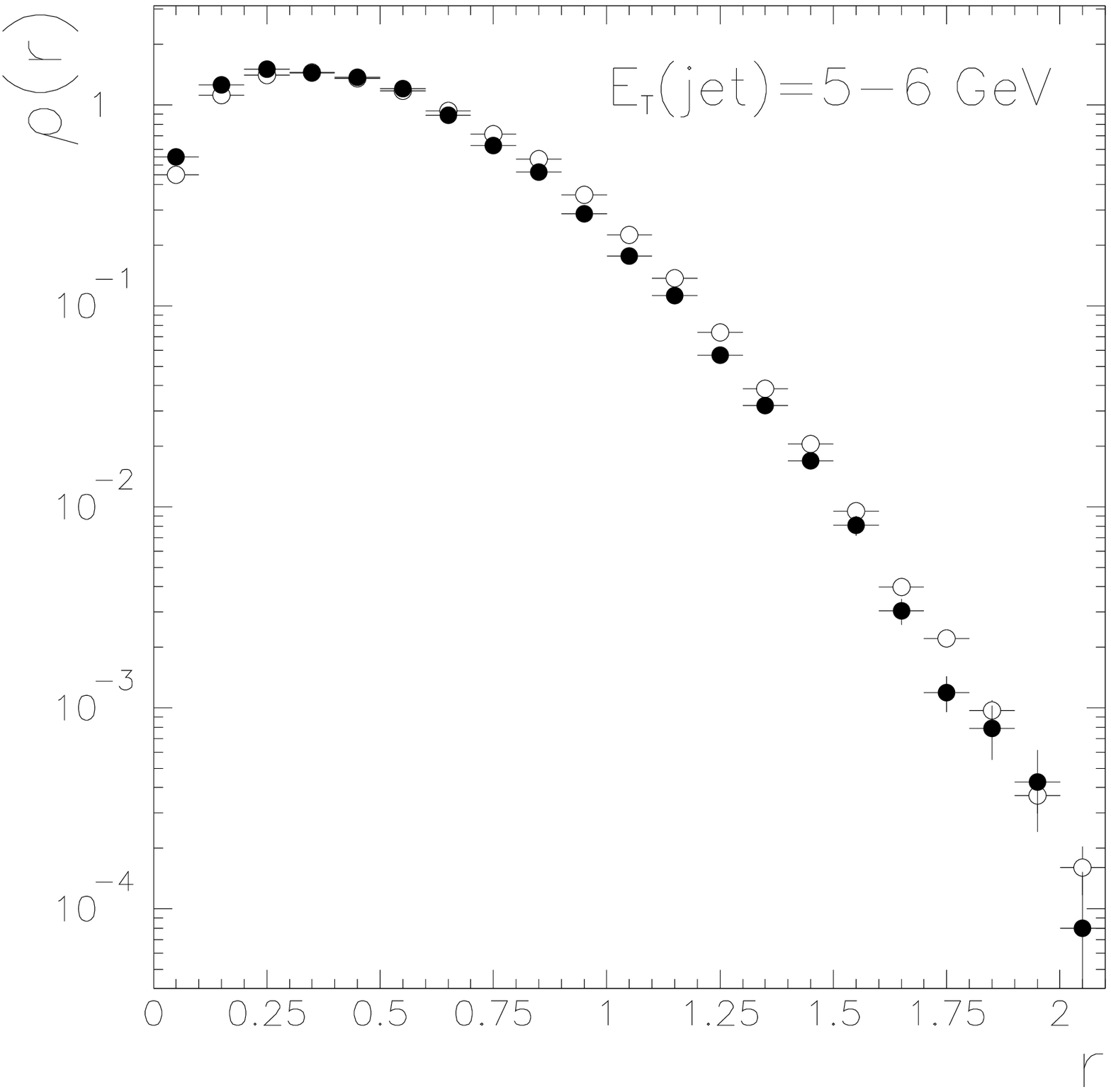,height=5.0cm,width=5.0cm}%
\epsfig{figure=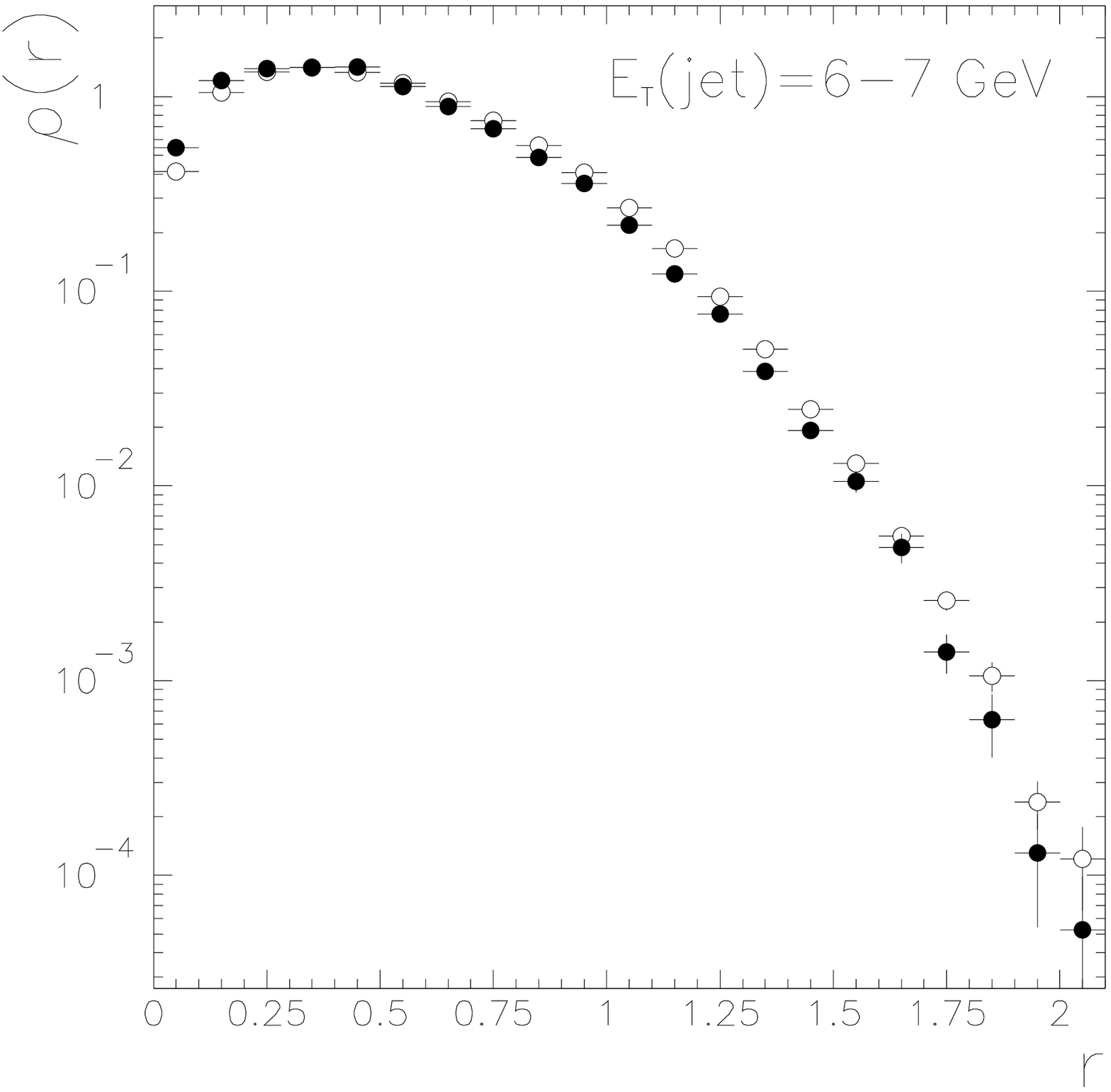,height=5.0cm,width=5.0cm}%
\epsfig{figure=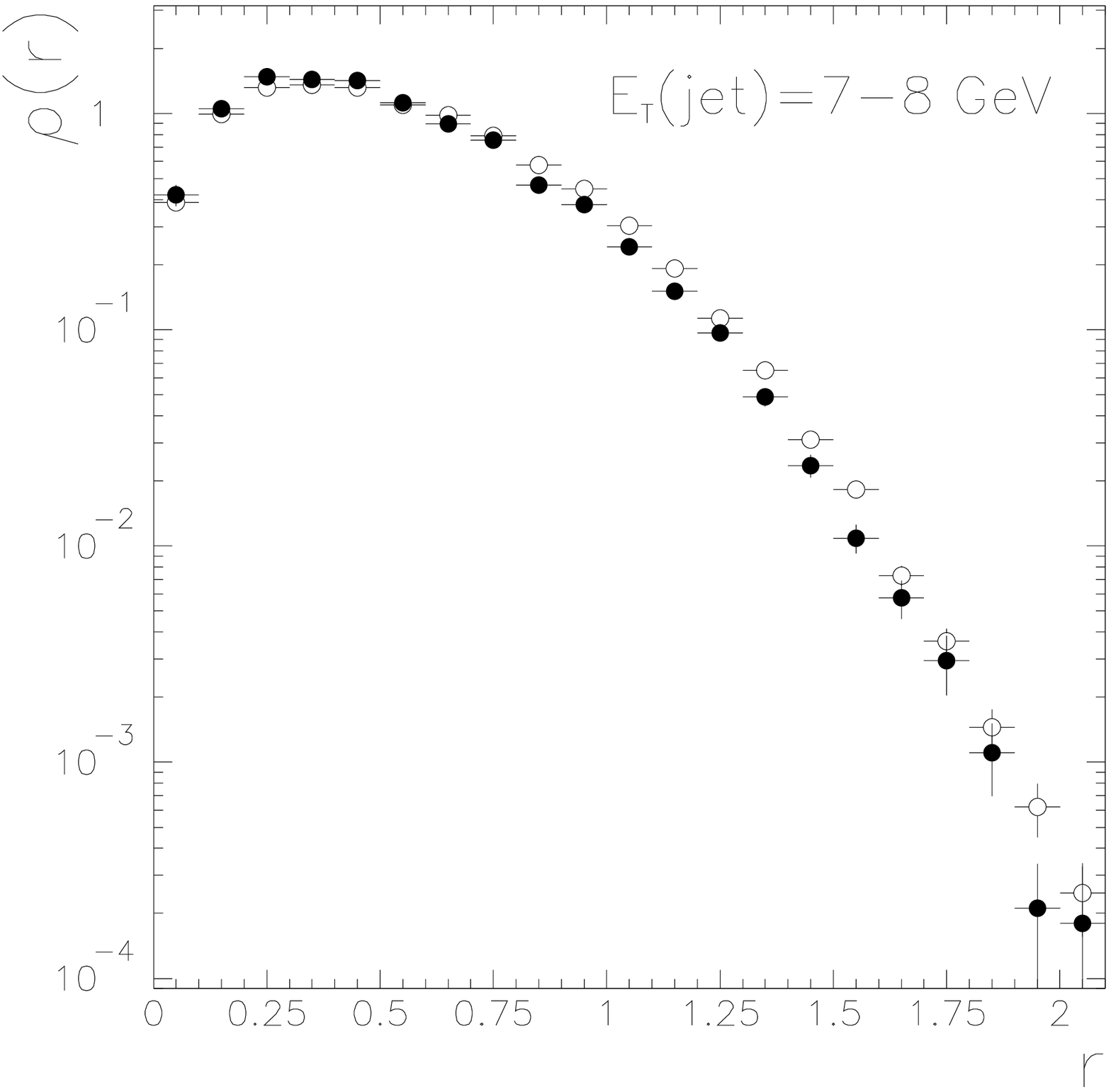,height=5.0cm,width=5.0cm}%
}%
\hbox{%
\epsfig{figure=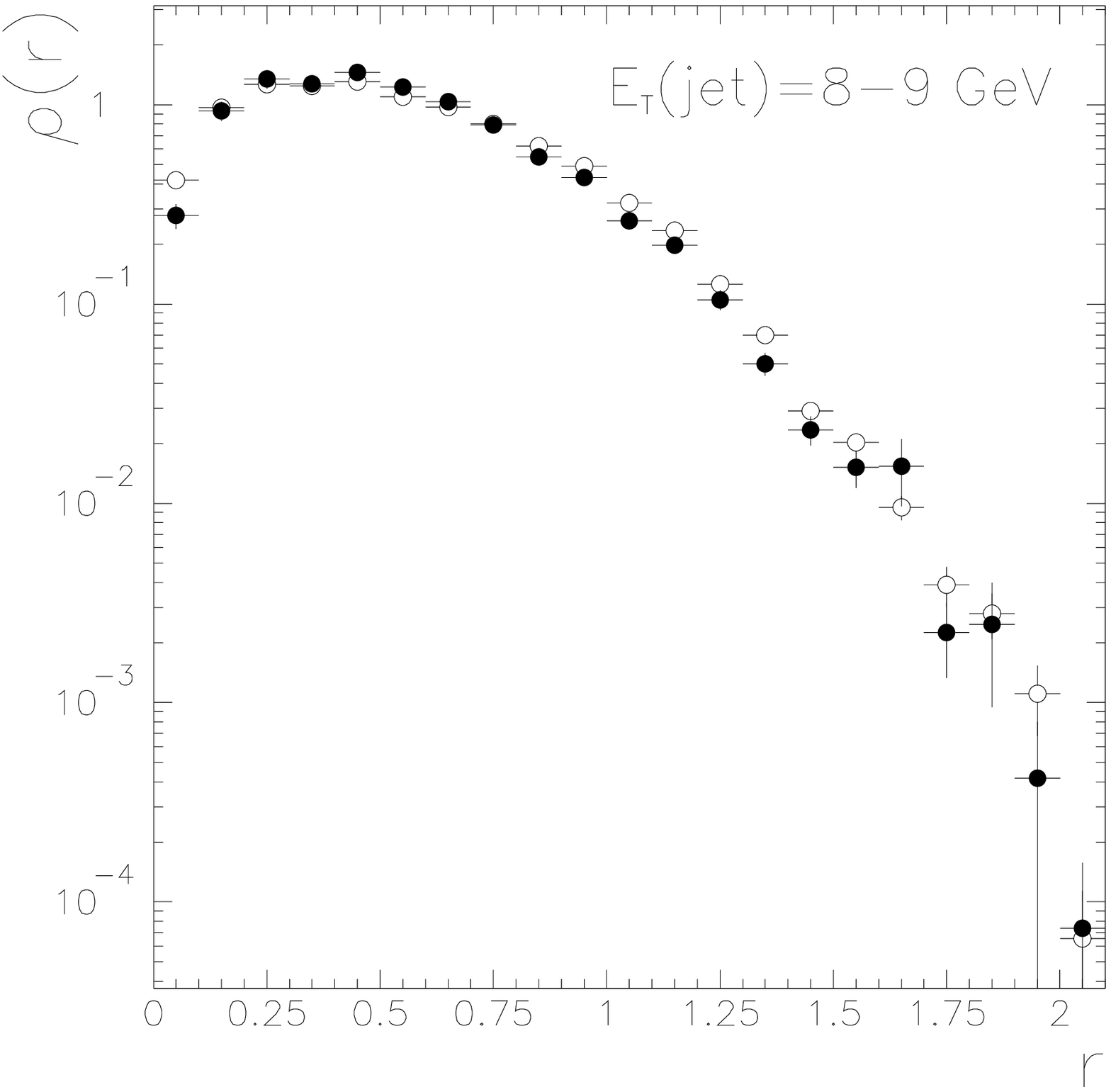,height=5.0cm,width=5.0cm}%
\epsfig{figure=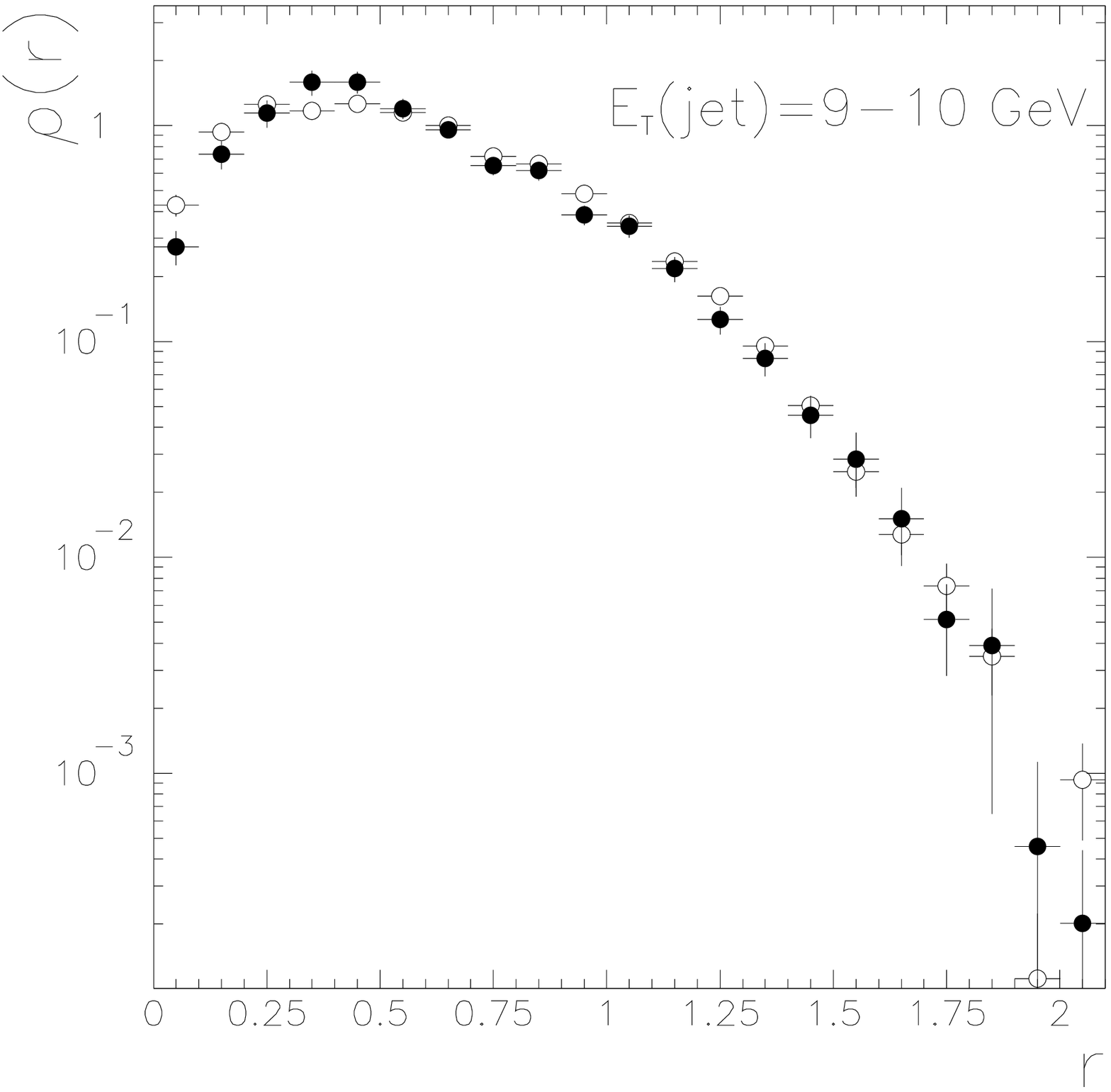,height=5.0cm,width=5.0cm}%
\epsfig{figure=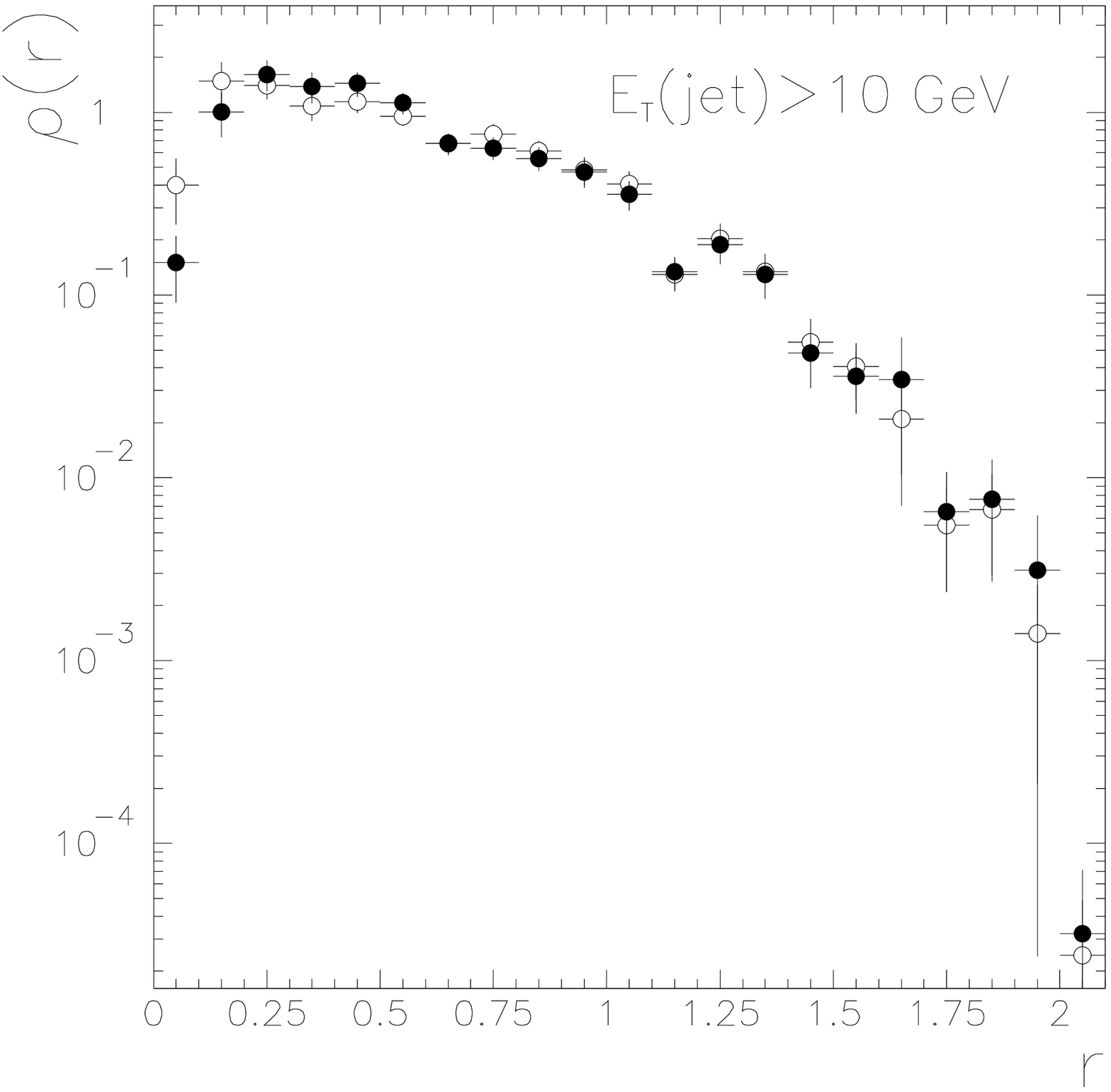,height=5.0cm,width=5.0cm}%
}%
}%
\end{center}
  \caption{\label{fig:rhos}
Comparison of the differential jet shape $\rho(r)$
(closed circles) with PYTHIA/FRITIOF predictions (open circles)
for the carbon target in every $E_T(jet)$ bin.
}
\end{figure*}

% ===========================================================
Fig.\ref{fig:rhos} shows the comparison of the
differential jet shape $\rho(r)$~ for carbon
target for six bins of jet transverse energy (closed circles)
and PY\-THIA/\-FRI\-TIOF predictions (open circles). 
The agreement between data and model predictions is quite good.

% ===========================================================
%\nextpage

% ===========================================================

\begin{figure*}[htb]
%\centerline{\psfig{figure=ratios.eps,height=9.0cm,width=16.0cm}}%
\begin{center}
\vbox{%
\hbox{%
\epsfig{figure=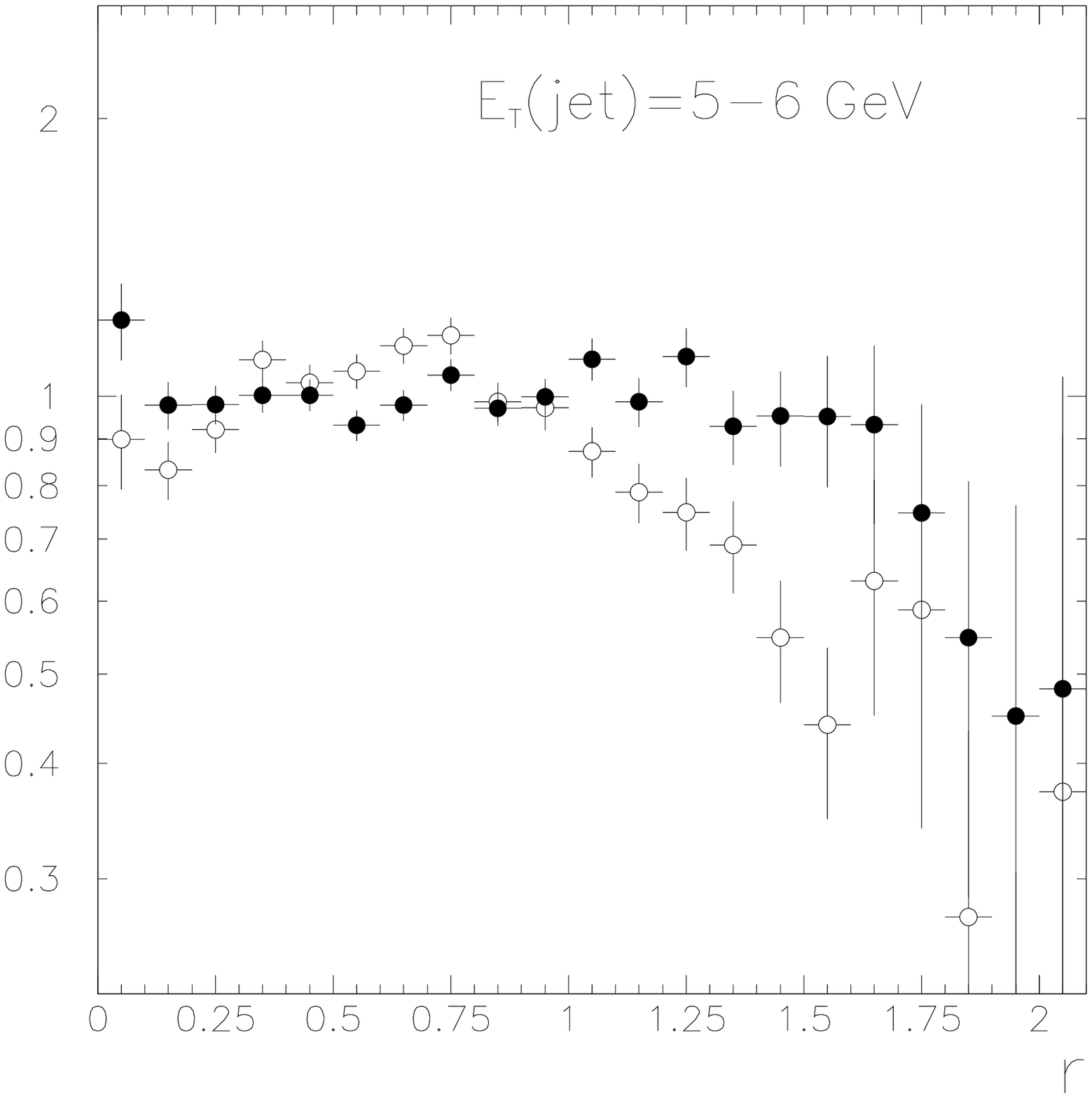,height=5.0cm,width=5.0cm}%
\epsfig{figure=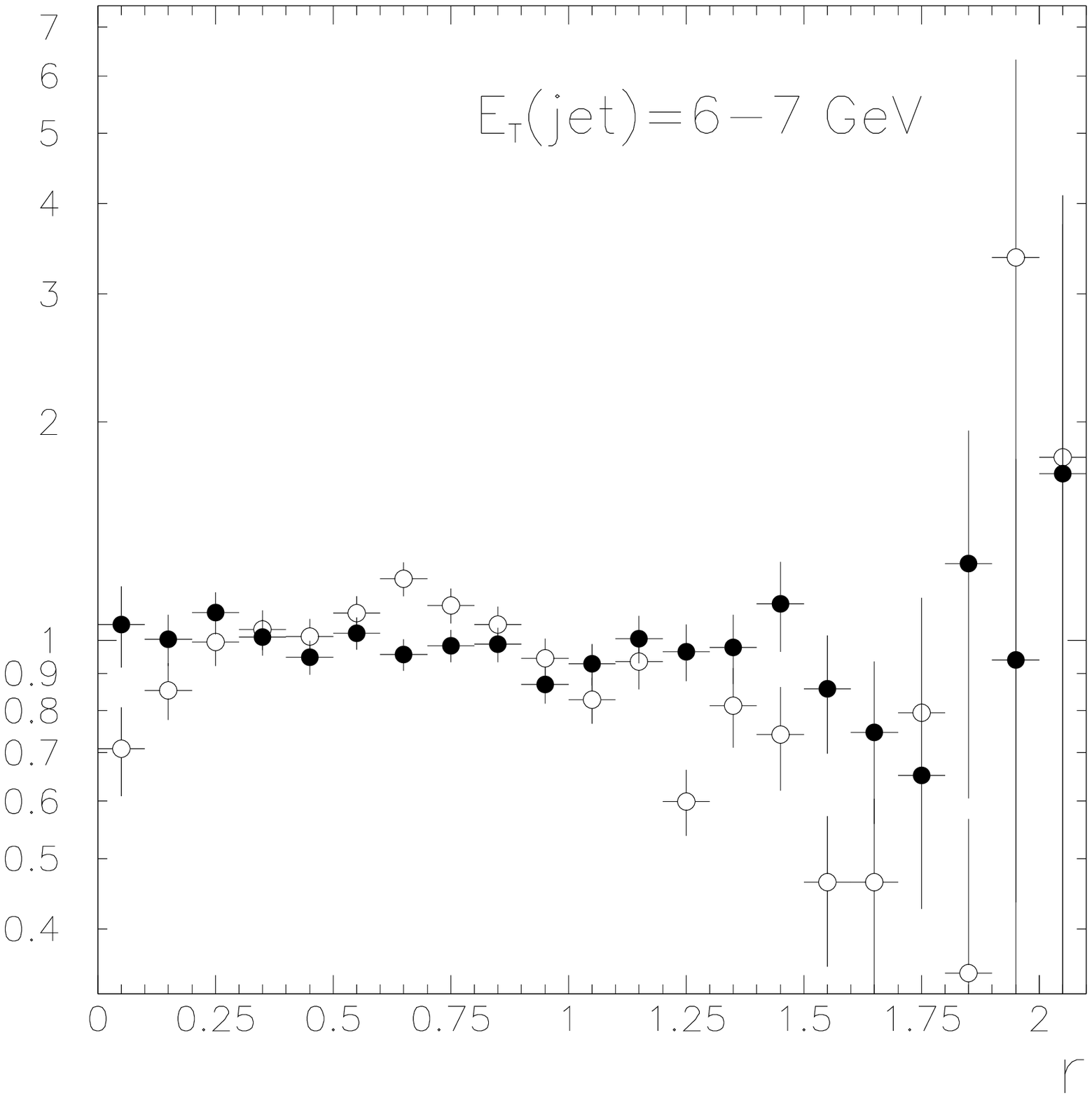,height=5.0cm,width=5.0cm}%
\epsfig{figure=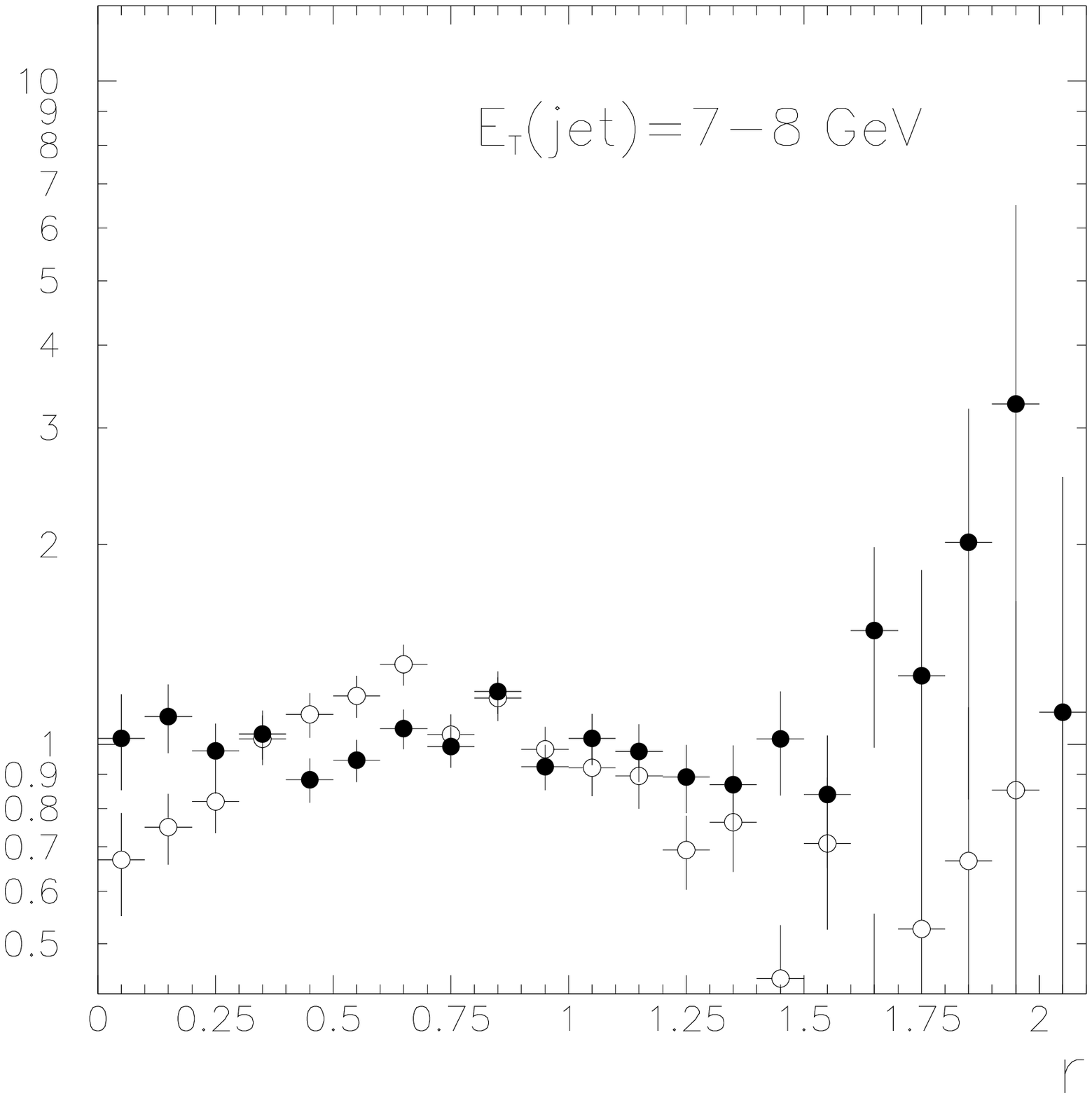,height=5.0cm,width=5.0cm}%
}%
\hbox{%
\epsfig{figure=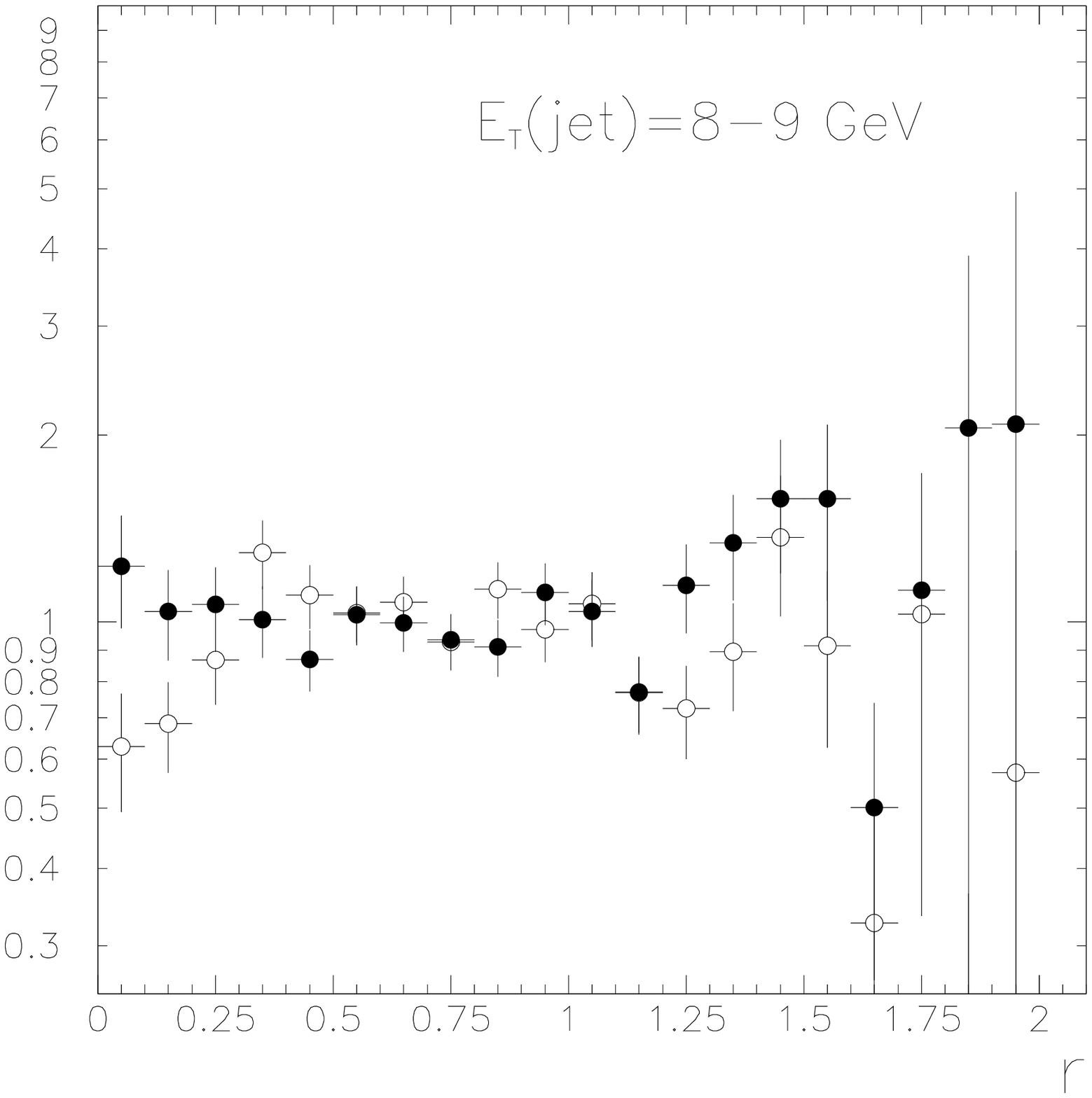,height=5.0cm,width=5.0cm}%
\epsfig{figure=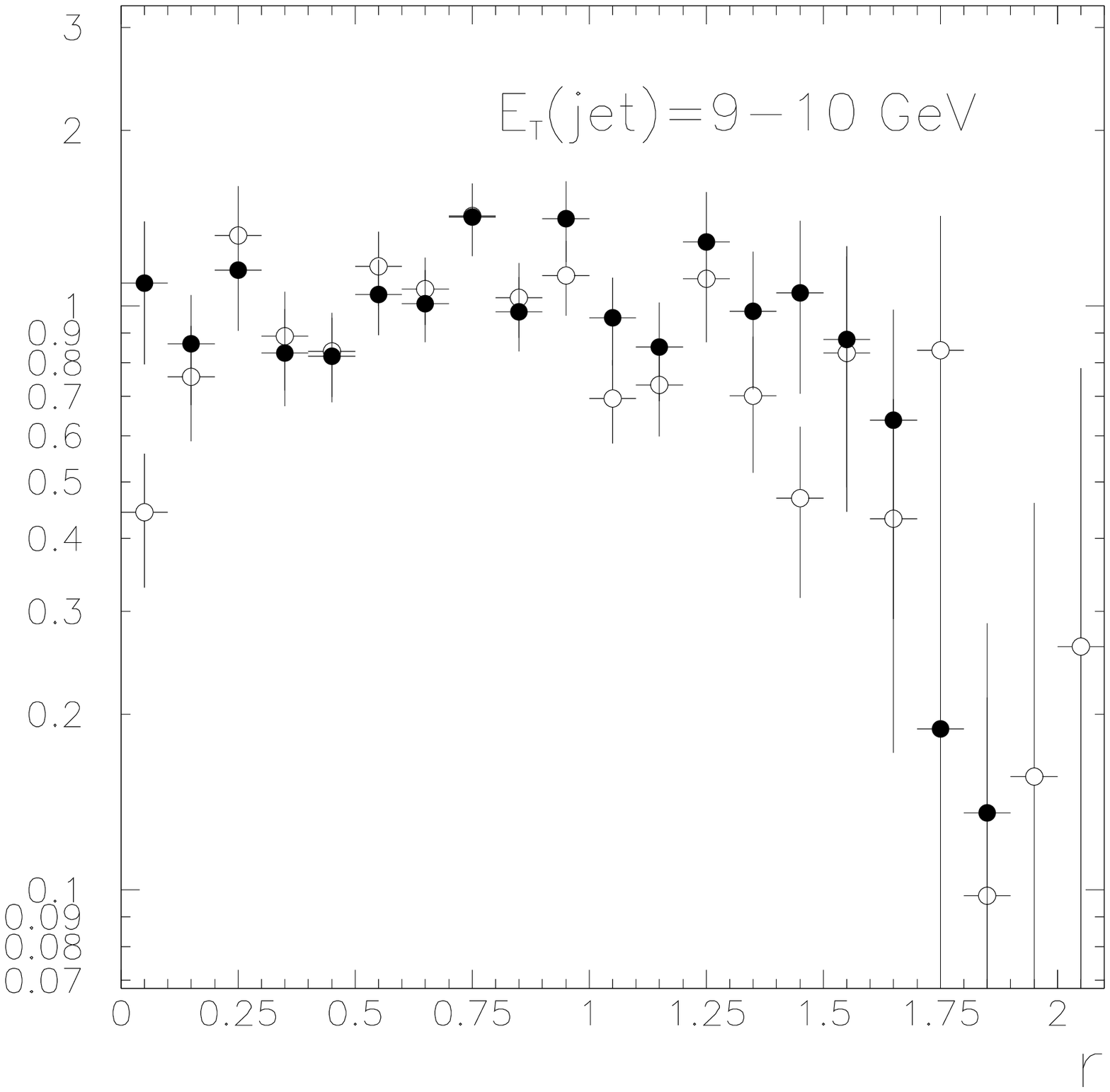,height=5.0cm,width=5.0cm}%
\epsfig{figure=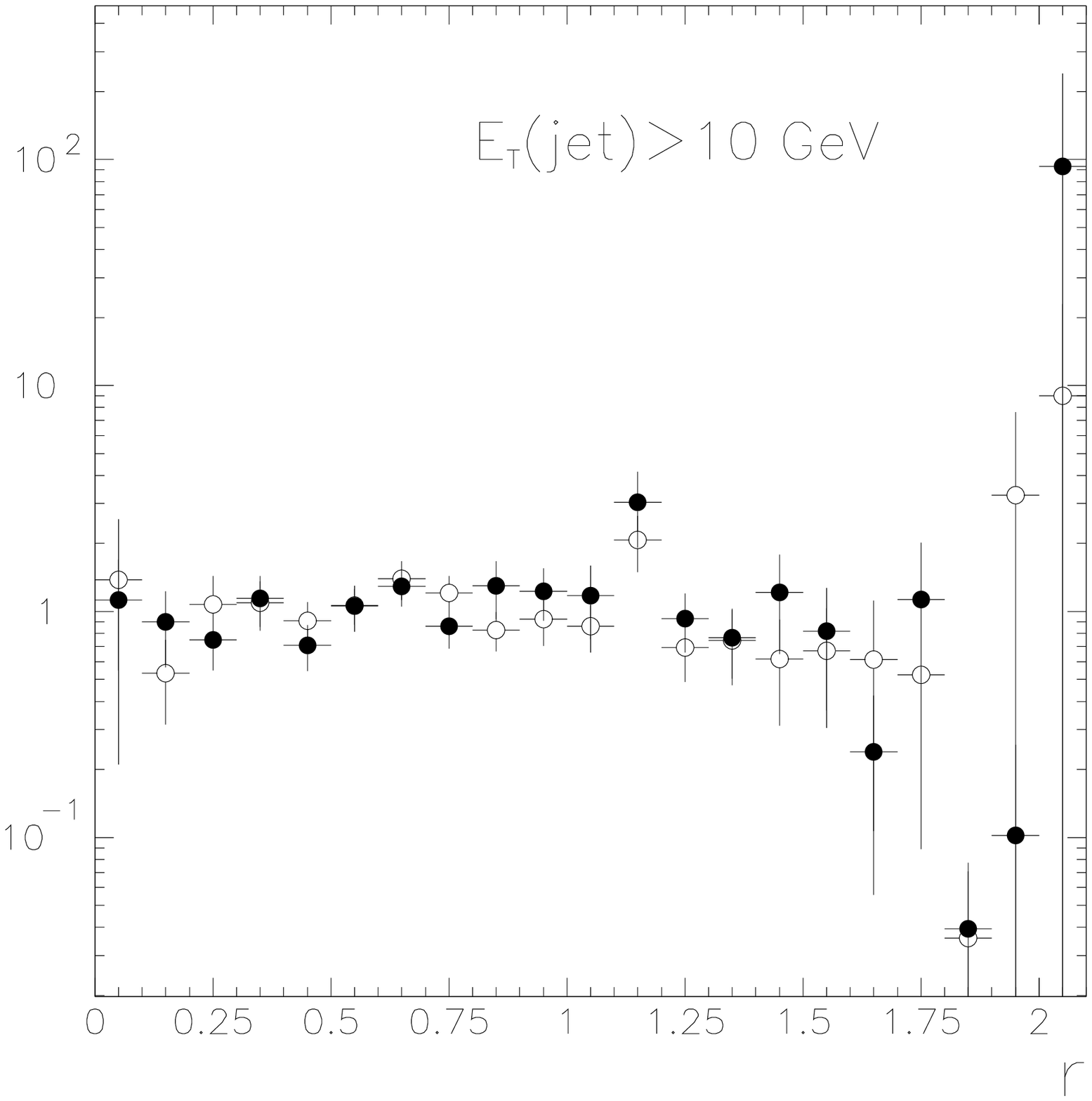,height=5.0cm,width=5.0cm}%
}%
}%
\end{center}
  \caption{\label{fig:ratios}
Ratios of the $\rho(r)$~ measured with aluminium (closed circles) 
and tungsten (open circles) targets to carbon results in every
$E_T(jet)$ bin.
}
\end{figure*}

Fig.\ref{fig:ratios} presents the ratios of the $\rho(r)$~ measured
with aluminium (closed circles) and tungsten (open circles) 
targets to carbon results. The results for aluminium and carbon targets
are nearly the same (ratio is approximately equal to unity).
Statistically significant deviations from unity exist for the tungsten 
sample at $r\,\approx\,0$ in $E_T$ range $5\,<\,E_T\,<\,10$~ GeV
and for large radius, $r\,>\,1$ in the bin $5\,<\,E_T\,<\,6$~ GeV. 
Thus, on average, jets produced on tungsten
nuclei have slightly different structure than jets for light nuclei.
However the position of maximum energy flow is the same for all nuclear 
targets, $r_{max}\,\approx\,0.3$.

% ===========================================================
%\nextpage

% ===========================================================

A jet is not a point-like object and has finite transverse size,
its radius. In the cone algorithm, the jet radius is fixed and equal 
to the size of the cone chosen for jet selection, $R_{cone}\,=\,0.7-1.0$.
In case of the $k_T$~algorithm, the jet radius is not fixed and extends 
to values $r\,>\,1$.
In the case of restricted acceptance over pseudo-rapidity 
(in the $\phi$-direction the acceptance is not restricted) 
one can encounter problems with the selection
efficiency of wide jets, when the jet radius exceeds the detector acceptance.
We performed measurements for jets with axes within the pseudo-rapidities 
$\eta(jet)\,=\,3.5\pm 0.1$. 
A realistic acceptance of HERA-B tracker (probability to register a track 
is greater than 10\%) covers the pseudo-rapidity range
$2.2\,<\,\eta\,<\,4.8$~ (calorimeter has larger acceptance). Therefore,
some fraction of selected jets must have particles outside 
the acceptance of our detector. 
However, the integral jet shape $\Psi(r)$ is about unity
already for $r\,\approx\,1$, i.e., almost all of the jet energy flow 
is concentrated within this cone and the whole jet is 
within the detector acceptance $(2.2\,<\,\eta\,<\,4.8)$.

\begin{figure}[htb]
\centerline{\epsfig{figure=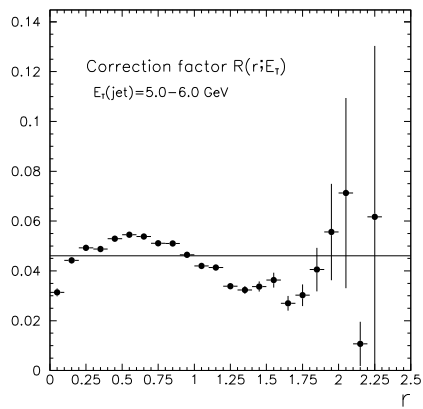,height=6.0cm,width=6.0cm}}%
  \caption{\label{fig:corrs_car}
The dependence of the systematic correction factor $R(r;E_T)$
on the distance to the jet axis for carbon target in 
$E_T(jet)\,=\,5.0-6.0$~GeV bin. The correction factor
is normalized to unity.
}
\end{figure}

Fig.\ref{fig:corrs_car} demonstrates the correction factor $R(r;E_T)$
from (\ref{corrs}) normalized to unity for the carbon target 
in the bin $E_T(jet)\,=\,5.0-6.0$~GeV.
The horizontal line is the mean value for this distribution.
The variation of the correction factor does not exceed 
$\approx\,25\%$ with respect to its mean value at radii $r\,\le\,1.5$
with good statistical errors. Thus we can expect that our systematic
corrections are reliable within the effective interval of jet radii. 
Note here that correction factors for different nuclear targets
agree well within their errors. Therefore they cancel out in the ratios 
of $\rho(r)$ for different nuclei, presented in Fig.\ref{fig:ratios}.
 
% ===========================================================
%\nextpage

% ===========================================================

% ===========================================================
\section{Conclusion}
\label{sec:concl}

We performed measurements of the differential $\rho(r)$ and integral $\Psi(r)$
jet shapes (\ref{jshape}) in the range of jet transverse energies 
$5\,<\,E_T(jet)\,<\,14$~GeV in proton--nucleus collisions at a proton
momentum 920~GeV/c. For jet selection, we used the longitudinally invariant
$k_T$~algorithm, defining jets according to standard Snowmass convention. 
We performed these measurements for three target nuclei --- carbon, aluminium 
and tungsten. For systematic corrections to the differential jet shape we used 
bin-by-bin statistical corrections, based on a Monte Carlo simulation 
of jet production, in each bin of jet transverse energy. Integral
jet shapes have been obtained by integration of the differential jet shapes.

We find good agreement between data and the predictions
of the PY\-THIA~6.2/\-FRITIOF~7.02 model with intrinsic transverse momentum
of partons $k_T\,=\,1$~GeV and independent frag\-mentation of partons
according to the symmetric Lund scheme.
The had\-ro\-ni\-za\-tion scheme in PYTHIA~6.2 at studied jet transverse
energies does not include showering of the scattered parton. 
It instead produces the final state hadrons,
mostly vector mesons, directly from the scattered parton. 
The agreement between our measurements and simulation shows
that in the considered transverse energy range $E_T(jet)\,<\,14$~ GeV, 
the parton cascade is still not developed and the hadronization has mainly
a non-perturbative nature, i.e., direct transition $parton\,\to\,hadrons$
and can be well reproduced by PYTHIA~6.2 with default parameters and
independent fragmentation of outgoing partons.

From comparison of the differential jet shape $\rho(r)$ for different nuclei,
we can conclude that the differences
in jet properties for carbon and aluminium targets are small.
Nevertheless, the transverse energy flow at small and large radii 
for the tungsten sample is slightly less than for light nuclei.
Note here that this observation does not depend on our MC model due to
cancellation of the systematic corrections in the ratios of $\rho(r)$
for different nuclei. 
This observation indicates some influence of the nuclear environment
on the formation of jets in heavy nuclei, especially at lower transverse 
energies, $5\,<\,E_T(jet)\,<\,6$~GeV. Possibly such a ''broadening''
of jets produced on a heavy nucleus occurs due to re-scattering of the jet
hadrons on nucleons which effect must be small for light nuclei.

% ===========================================================
%\centerline{\large\bf Acknowledgments}
\acknowledgement{
We express our gratitude to the DESY laboratory for the
financial and technical support during this work.
We are also indebted to the HERA-B Collaboration for permission 
to use the experimental data and for fruitful cooperation.
Our special thanks to Spokesman of the HERA-B Collaboration 
Dr.~M.~Medinnis for numerous useful and stimulating discussions.
We want also to thank all the HERA-B members who contributed
in discussing and improving of the manuscript.
}

% ===========================================================
%\newpage

%  ------------- References -----------------------

% - - - - - - - - - - - - - - - - - - - - - - - - - - - - - -

\end{document}